\documentclass[aps,prb,twocolumn,superscriptaddress,showpacs,floatfix]{revtex4-2}
\usepackage{graphicx}
\usepackage{amsmath,amssymb}
\usepackage{bm}
\usepackage{xcolor}
\usepackage{hyperref}
\hypersetup{colorlinks=true,linkcolor=blue,citecolor=blue,urlcolor=blue}

\newcommand{\sigl}{\sigma_{\ell}}
\newcommand{\sigH}{\sigma_{\mathrm{H}}}
\newcommand{\mus}{\mu^{\star}}

\newcommand{\um}{\,\mu\mathrm{m}}

\begin{document}

\title{Chiral or unidirectional? Multiterminal current partition in the quantum Hall regime from self-consistent screening theory}

\author{Afif Siddiki}
\email{afif.siddiki@atlas.edu.tr}
\affiliation{Atlas University Vocational School, \.{I}stanbul, Turkey}

\date{\today}

\begin{abstract}
Multiterminal current measurements on a six-terminal Hall bar were recently used to argue that the current flow is chiral on the plateaus of the quantized Hall resistance and non-chiral between them, and, in a companion analysis, that the imposed current flows along one edge only. We examine these conclusions with the self-consistent screening theory of the quantum Hall effect. We first show that the plateau values of the current partition follow from a vanishing longitudinal conductivity and a quantized Hall conductivity alone: every two-dimensional conductor with $\sigl=0$ and $\sigH=\nu e^2/h$ has the conductance matrix of the ``coherent and chiral'' model, whatever the spatial distribution of the current. More generally, terminal currents determine the current density only up to circulating currents, a known indeterminacy that we make explicit with a stream function, and the chirality of the current density is a property of exactly this part. A two-dimensional local transport calculation for the measured device, including its two interior contacts, with the self-consistent profiles mapped into the device geometry, reproduces the plateau partition for all contact configurations and the ordering of the configuration-dependent suppression between the plateaus without edge channels and without parameters adjusted within the calculation, the sample parameters and contact resistances being those of the experiment; a three-number phenomenology for long-range disorder and spin splitting then recovers the measured plateau width, longitudinal resistance and suppression magnitude. Each observation and each explanation of the two papers, including the fitted mobility between the plateaus and the finite-current statements about persistent and imposed currents, has a counterpart in this solution, quantitative on the plateaus and semi-quantitative between them. The self-consistent current density separates into an equilibrium (magnetization) current, which is chiral and carries no net current, and a transport current driven by the gradient of the electrochemical potential, which, in the cross sections between the contacts, flows in the same direction in both edge strips and in the bulk strip. The ``one-edge'' imposed current is a bookkeeping choice of the reference chemical potential; the bias-induced current is shared between both edges, as observed by scanning-force microscopy. The solution predicts the potentials of the floating interior contacts, read with an input impedance above their isolation resistance: equal when the bulk around them is quantized, and differing by up to half the Hall voltage when the current flows in edge strips around a compressible bulk. Multiterminal current partition measures the sign of the Hall conductivity and the absence of dissipation, not the location or the chirality of the transport current. The agreement with the chiral model shows that the terminal currents are those of the dissipationless limit and bounds the dissipation the contacts can see; the edge channels are the premise of the interpretation, not a result of the measurement.
\end{abstract}

\maketitle

\section{Introduction}

Where the current flows in a Hall bar under the conditions of the integer quantized Hall effect (QHE) has been debated since the discovery of the effect~\cite{vonKlitzing1980,Halperin1982,MacDonald1983,Buettiker1988,Thouless1993,Komiyama1996,Weis2011}. Two limiting pictures are commonly contrasted. In the Landauer-B\"uttiker picture~\cite{Buettiker1988,Haug1993} the current is carried by one-dimensional chiral edge channels, each contributing $e^2/h$, and the quantization follows from the absence of backscattering; experiments with selectively populated and probed edge channels~\cite{vanWees1989,Komiyama1989,Alphenaar1990} established the picture for the terminal currents. In the screening picture~\cite{Chklovskii1992,Lier1994,Guven2003,Siddiki2004,Gerhardts2008} the electron density of the interacting two-dimensional electron system (2DES) arranges itself into compressible regions, where the Landau-level (LL) energy is pinned to the chemical potential, and incompressible strips (ICSs) of integer filling, where it is not; the Hall voltage drops across the ICSs and the dissipationless current flows there. Scanning-force-microscopy (SFM) measurements of the Hall potential landscape~\cite{Weitz2000,Ahlswede2001,Ahlswede2002,Dahlem2010,Panos2014}, together with electro-optic~\cite{Fontein1991}, scanning-capacitance~\cite{Suddards2012}, microwave-impedance~\cite{Lai2011} and scanning-gate~\cite{Pascher2014} imaging, have shown the second picture at work: on the low-field side of a plateau the Hall voltage drops near both edges, in the plateau center it drops across the bulk, and between plateaus it drops linearly across the sample.

Recently Sirt, Kamm, Umansky and Ludwig~\cite{Sirt2025NJP} measured the partition of the current emitted by one contact of a six-terminal Hall bar into the other, grounded contacts, as a function of the magnetic field $B$ and of the configuration of grounded contacts. They compared the measured current ratios with two limiting models, a coherent chiral conductance matrix and a multiterminal Drude model, and concluded that the current flow is chiral on the plateaus and non-chiral between them. In a companion analysis~\cite{Sirt2024arXiv} they decomposed the current density in a Hall bar into a persistent and an imposed part and concluded that the imposed current flows entirely along the edge with the higher chemical potential, in explicit contrast to the local-equilibrium calculations of G\"uven and Gerhardts~\cite{Guven2003}, in which the imposed current flows in the same direction in both edge strips.

The experiment is clean and the data are valuable. The purpose of this paper is to ask, with the same tools that were used to establish the screening picture, what such a measurement decides and what it does not. The answer will be that everything it decides is reproduced by a model in which the current flows in one direction only, and that the question it was meant to settle can be settled by the two interior contacts of the same sample. We proceed in four steps. (i)~We show that the plateau values of the multiterminal current ratios follow from $\sigl=0$ and $\sigH=\nu e^2/h$ alone, for any spatial distribution of the current, and, more generally, that terminal currents determine the current density only up to circulating currents, so that its chirality lies beyond their reach. That resistances do not determine the current distribution has been remarked before~\cite{Beenakker1991,Komiyama1996,SchulzBaldes2000}; what is new here is the explicit form of the indeterminacy and its application to the device and the claims of Refs.~\cite{Sirt2025NJP,Sirt2024arXiv}. (ii)~We solve the self-consistent Thomas--Fermi--Poisson problem for the parameters of the measured device, compute the local conductivity tensor, and solve the two-dimensional local transport problem for the actual six-terminal geometry and the measured contact resistances. This reproduces the plateau partition for all configurations and the configuration-dependent suppression between the plateaus. (iii)~We separate the self-consistent current density into its equilibrium and transport parts. The equilibrium (magnetization) current is chiral and carries no net current; the transport current, which carries the measured current and the dissipation, flows in the same direction in both edge strips. (iv)~We compute the potentials of the floating contacts, among them the two interior contacts of the sample, and the four-terminal resistances; the interior contacts provide the test of the current distribution that the terminal currents cannot. The question in the title therefore has a two-part answer: the equilibrium current is chiral, the transport current is unidirectional, and the multiterminal partition is blind to the difference.

Throughout we use ``chiral'' in the sense of Refs.~\cite{Sirt2025NJP,Sirt2024arXiv}: current flowing in opposite directions along the two edges of the bar. We use ``unidirectional'' for a current density that has the same sign across a cross section of the bar. In the broader sense of a direction fixed by the sign of $B$, the local-Ohm solution is chiral as well: the electrochemical potential propagates one way around the boundary, and the terminal response is directional. The question of this paper is not whether the quantum Hall effect is chiral in that sense, which it is in every description, but whether the current density counter-flows at the two edges, and whether terminal currents can tell.

\section{What the current partition measures}
\label{sec:formal}

Consider a 2DES of arbitrary shape with $n$ ideal contacts along its boundary and a local conductivity tensor. In steady state the transport current density is $\bm{j}=\hat\sigma\,\bm{\nabla}\mus/e$, where $\mus(\bm r)$ is the electrochemical potential, $\bm\nabla\cdot\bm j=0$, and $\bm j\cdot\bm n=0$ on the free (etched) parts of the boundary. With
\begin{equation}
\hat\sigma=\begin{pmatrix}\sigl & \sigH\\ -\sigH & \sigl\end{pmatrix},
\label{eq:sigma}
\end{equation}
the normal component of the current at a free boundary with tangent $\bm t$ is $\bm j\cdot\bm n=(\sigl\,\partial_n\mus+\sigH\,\partial_t\mus)/e$. Let now $\sigl=0$ and $\sigH=\nu e^2/h$ throughout the sample, the idealization of a plateau. The condition $\bm j\cdot\bm n=0$ then requires $\partial_t\mus=0$: \emph{every free boundary segment between two contacts is an equipotential}. The current through any line connecting a point $A$ on one free segment to a point $B$ on another is
\begin{equation}
I_{AB}=\int_A^B\bm j\cdot\bm n_c\,ds=\frac{\sigH}{e}\left[\mus(B)-\mus(A)\right],
\label{eq:IAB}
\end{equation}
since $\bm j\cdot\bm n_c=\sigH\,\partial_s\mus/e$ along the line; the result is independent of the path and of how $\bm j$ is distributed along it. (In the screening picture $\sigH$ varies and $\sigl$ is finite inside the compressible regions, but on a plateau these regions are equipotential pockets attached to the contacts, so that $\partial_s\mus\neq0$ only inside the strips, where $\sigH=\nu e^2/h$; Sec.~\ref{sec:partition} shows how closely the self-consistent solution approaches the ideal case.) Number the contacts $1,\dots,n$ along the boundary and let $\mus_{k,k+1}$ be the potential of the free segment between contacts $k$ and $k+1$. Current conservation at contact $k$ gives
\begin{equation}
I_k=\frac{\sigH}{e}\left(\mus_{k-1,k}-\mus_{k,k+1}\right).
\label{eq:Ik}
\end{equation}
Finally, in the limit $\sigl\to0^+$ the electrochemical potential of the free segment between $k$ and $k+1$ equals that of one of its two end contacts, and which one is fixed by the sign of $\sigH$, i.e.\ by the direction of $B$. Three conditions enter here and should be stated. The limit is $\sigl\to0^+$, not $\sigl=0$: at $\sigl=0$ exactly, $\bm\nabla\cdot(\hat\sigma\bm\nabla\mus)$ vanishes identically for uniform $\sigH$ and the interior potential is undetermined, which is the same indeterminacy from the other side; a small residual $\sigl$ makes $\mus$ harmonic in the interior and selects the hot-spot corners by the sign of $\sigH$. Equation~(\ref{eq:IAB}) requires $\sigl=0$ along the cut, or more generally that every point of the cut lies either in a dissipationless region or in an equipotential pocket, where both terms of $\bm j\cdot\bm n$ vanish; a dissipative region with a potential gradient across the cut breaks it, which is the regime between the plateaus. And interior contacts add holes to the domain, around which circulating currents are allowed and through which the floating condition fixes only the net current; Sec.~\ref{sec:probes} returns to them. This is the classical ``hot-spot'' solution of the Hall problem at vanishing dissipation~\cite{Wakabayashi1978,Rendell1981,Klass1991}: the potential jumps across the sample along a line connecting two diagonally opposite contact corners, and each free edge segment carries the potential of the contact at its upstream end. With $\mus_{k,k+1}=\mus_k$ Eq.~(\ref{eq:Ik}) becomes
\begin{equation}
I_k=\frac{\sigH}{e}\left(\mus_{k-1}-\mus_k\right),\qquad G_{k,k-1}=\sigH=\nu\frac{e^2}{h},
\label{eq:LB}
\end{equation}
which is the conductance matrix of Eq.~(3) of Ref.~\cite{Sirt2025NJP} and, with the contact resistances $R_k$ added in series, their Eq.~(4).

Three remarks follow. First, Eq.~(\ref{eq:LB}) holds for \emph{any} two-dimensional conductor with $\sigl=0$ and $\sigH=\nu e^2/h$, whether the current flows in one-dimensional edge channels, in incompressible strips of finite width, in the bulk, or in any combination. The terminal currents fix only the differences of the boundary potentials, Eq.~(\ref{eq:IAB}); the interior distribution of $\bm j$ drops out. This is the content of Thouless's remark that edge voltages do not locate the current~\cite{Thouless1993} and of the two-representation analysis of Komiyama and Hirai~\cite{Komiyama1996}. The authors of Ref.~\cite{Sirt2025NJP} note the same equivalence themselves when they observe (their Sec.~II\,B and Appendix~B) that the Hall part of their Drude model, which they describe as unidirectional and ``in no way chiral'', is ``mathematically equivalent'' to the chiral model for $\mu B\to\infty$. Second, the only place where a chirality enters is the assignment $\mus_{k,k+1}=\mus_k$ rather than $\mus_{k+1}$, and it is fixed by the sign of the Hall conductivity. What the plateau values of the current ratios establish is therefore $\sigl=0$ together with the sign and the quantized magnitude of $\sigH$. Third, a finite $\sigl$ breaks Eq.~(\ref{eq:IAB}): free boundary segments are no longer equipotentials, part of the current reaches contacts by dissipative paths, and the partition depends on the geometry and on which contacts are grounded. The configuration dependence observed between the plateaus~\cite{Sirt2025NJP} is the signature of a finite local $\sigl$, which both the Drude model and the screening theory contain.

\begin{figure*}
\includegraphics[width=\textwidth]{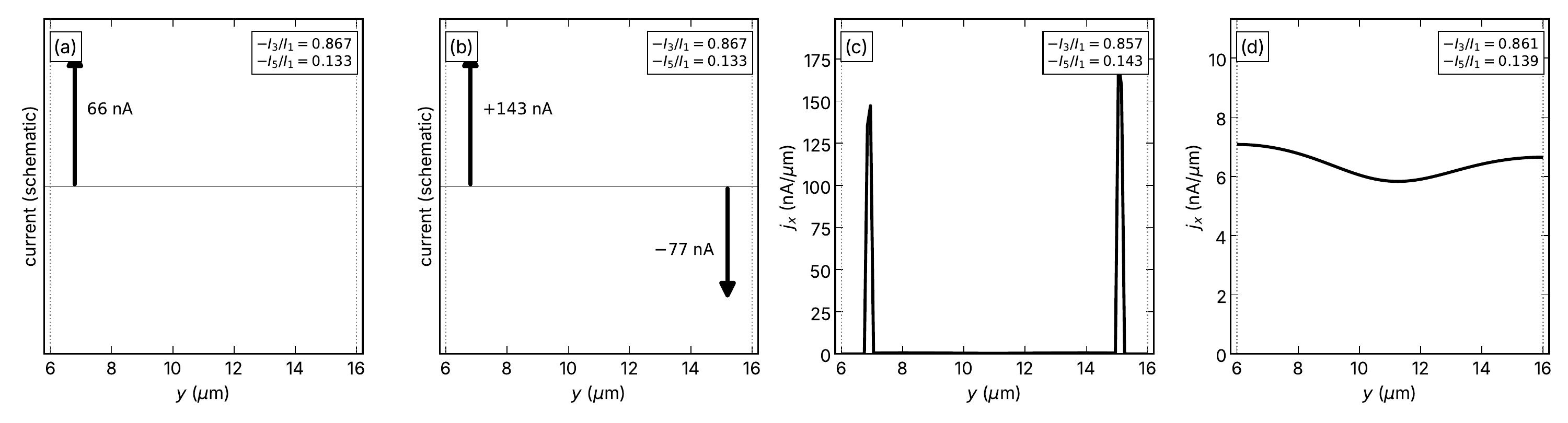}
\caption{Four current densities across the bar of Ref.~\cite{Sirt2025NJP} ($x=20\um$, between the arms, 3T configuration, $\nu=2$); (a)--(c) produce the same terminal currents, and (d) approaches them as $\sigl\to0$. (a)~The current confined to the high-potential edge, the ``imposed current'' of Ref.~\cite{Sirt2024arXiv}. (b)~Counter-flowing edge currents, the Landauer--B\"uttiker picture including the equilibrium currents, drawn with the complete equilibrium current per edge of Sec.~\ref{sec:currents} ($M_{\rm bulk}=110$~nA at $2.3$~T); any pair $I_{\rm h}$, $-I_{\rm l}$ with $I_{\rm h}-I_{\rm l}=I_1$ gives the same terminal currents. (c)~The transport current of the self-consistent screening theory, two-dimensional solution of Sec.~\ref{sec:partition} at $2.3$~T: two strips, same direction. (d)~A homogeneous conductor with $\sigH=2e^2/h$ and $\sigl=0.01\,e^2/h$, the Drude model of Ref.~\cite{Sirt2025NJP} with $\mu B=200$: the current is spread over the whole width. The boxes give the partition into the grounded contacts $3$ and $5$ with the measured contact resistances: in (a) and (b) by construction, Eq.~(\ref{eq:LB}); in (c) computed, the $1\%$ deficit coming from the weakly dissipative arm interiors of the clean model (Sec.~\ref{sec:partition}; with the localized bulk of Sec.~\ref{sec:model} the values are $0.867$ and $0.133$); in (d) computed, the deviation from the chiral values vanishing linearly with $\sigl$ ($0.749$, $0.818$, $0.851$ and $0.861$ for $\sigl=0.3$, $0.1$, $0.03$ and $0.01\,e^2/h$).}
\label{fig:equiv}
\end{figure*}

The remarks can be sharpened into a statement about what terminal currents can determine at all, independently of any conductivity model. The statement is not new in substance, resistance measurements being known not to determine the current distribution~\cite{Beenakker1991,Komiyama1996,SchulzBaldes2000}, but its explicit form is useful. Let $\bm j$ and $\bm j'$ be two steady current densities in the same sample with the same current $I_k$ at every contact. Their difference $\delta\bm j=\bm j'-\bm j$ is divergence-free, has no normal component on the free boundary, and has zero net flux through every contact. In two dimensions it derives from a stream function,
\begin{equation}
\delta\bm j=\hat{\bm z}\times\bm\nabla\psi ,
\label{eq:stream}
\end{equation}
and the flux of $\delta\bm j$ through any line from a point $A$ to a point $B$ is $\psi(B)-\psi(A)$. Along the boundary, $\psi$ is constant on each free segment, because $\delta\bm j\cdot\bm n=0$ there, and it returns to the same value at the two ends of every contact, because the net flux through the contact vanishes; it therefore takes one and the same value on all free segments of the outer boundary, and the flux of $\delta\bm j$ through any line connecting two points of the free boundary is zero. Two current densities with the same terminal currents thus carry the same net current through every cross section of the sample and differ only by a current that circulates in closed loops: around the sample, closing through the contacts, or around an interior contact. Conversely, any such circulating current can be added to $\bm j$ without changing a single terminal current. The construction is kinematic: it says which current densities the contacts cannot distinguish, not which of them a given Hamiltonian, electrostatics and set of contacts realize; the realized one is the model's, which is exactly why a statement about it is a statement about the model.

Chirality in the sense of Refs.~\cite{Sirt2025NJP,Sirt2024arXiv} is a property of precisely this circulating component. A current $I_{\rm h}$ along one edge and $-I_{\rm l}$ along the other, with $I_{\rm h}-I_{\rm l}=I$, differs from a current $I$ flowing in the same direction along both edges, or spread over the bulk, by a loop current $I_{\rm l}$ around the perimeter, and the terminal currents of the two are identical for any $I_{\rm l}$, in any geometry, with any number of contacts, at any field. Figure~\ref{fig:equiv} shows four such current densities for the device of Ref.~\cite{Sirt2025NJP}: the current confined to one edge, counter-flowing edge currents, the two same-direction strips of the screening theory, and the uniform flow of a homogeneous conductor with $\sigl\to0$; their partition into the grounded contacts is the same (for the homogeneous conductor in the limit). A multiterminal current measurement therefore determines the net current through every cross section and, through the assignment of the free-segment potentials, the sign of $\sigH$. Whether the current density also contains a counter-flowing part, and how large it is, it cannot tell: a statement about the chirality of the current density is an assumption of the model used to interpret the terminal currents, not a result of the measurement. Two qualifications. The null space contains every circulating current, bias-dependent ones included, such as the small loops around floating contacts in Sec.~\ref{sec:partition}; the equilibrium magnetization current is its physically realized member at $I=0$, not its definition. And the circulating part is unobservable by terminal measurements, not in principle: the equilibrium current is the orbital magnetization of the sample and is measured by torque or SQUID magnetometry~\cite{Eisenstein1985,Usher2009}.

This is where the chirality enters Ref.~\cite{Sirt2025NJP}. Its first model ``generically assumes coherent and chiral current flow'' and is written down as the conductance matrix (3); the measured partition on the plateaus agrees with it; and the agreement is then presented as a proof of chiral current flow ``if we accept that Eq.~(3) describes chiral current flow'' (their Sec.~III\,F). The condition carries the whole weight. With chiral flow defined by Eq.~(3) and Eq.~(3) confirmed by the data, the proof reduces to its definition: the test would be passed by every dissipationless Hall conductor, with or without edge channels. Equation~(3) of Ref.~\cite{Sirt2025NJP} is Eq.~(\ref{eq:LB}) here, the conductance matrix of every two-dimensional conductor with $\sigl=0$ and quantized $\sigH$, and, as noted above, Ref.~\cite{Sirt2025NJP} derives it a second time from a current that it describes as unidirectional and ``in no way chiral''. The adjective ``coherent'' carries no additional content either: a conductance matrix with transmission $\nu$ between adjacent contacts contains no phase, and the same matrix follows from the classical solution of the Hall problem at vanishing dissipation~\cite{Wakabayashi1978,Rendell1981,Klass1991}. What the measurement establishes is the matrix. The picture of a current confined to one edge between adjacent contacts is supplied by the Landauer--B\"uttiker derivation of the matrix, and by the preceding paragraph it could not have been established by terminal currents in any case. The reversal of the partition with the direction of $B$ (Fig.~6 of Ref.~\cite{Sirt2025NJP}), governed, as the authors note, by the direction of $B$ alone, is the reversal of the sign of $\sigH$; Sec.~\ref{sec:partition} shows that the unidirectional solution reverses in exactly the same way, with the current density at a cross section between the arms unchanged in sign. Reference~\cite{Sirt2025NJP} closes by stating that its result ``confirms the assumption of chiral current flow frequently used'' in the QHE. What it confirms is that the assumption is consistent with the data, which it would be for any dissipationless Hall conductor; the companion paper~\cite{Sirt2024arXiv} goes further and notes that ``the occurrence of bulk current is not in conflict with chiral current flow as long as scattering of the carriers is suppressed inside the bulk ICS'', at which point ``chiral'' has become a synonym for ``dissipationless'' and no longer says where the current flows.

The rest of the paper makes these statements quantitative for the measured device.

\section{Model}
\label{sec:model}

\subsection{Self-consistent density, potential and local conductivity}
\label{sec:1d}

The translation-invariant problem is solved with the program used in Refs.~\cite{Guven2003,Siddiki2004} (SCBLIN, here in a line-by-line Python translation of the original Fortran code), without changes to its physics. It treats a Hall bar of width $2d=10\um$ in the plane $z=0$ in the Thomas--Fermi--Poisson approximation~\cite{Siddiki2003,Siddiki2004}: the 2DES is confined by the field of a homogeneous donor layer of density $n_0=1.24\times10^{11}$~cm$^{-2}$, the Hartree potential is obtained from the two-dimensional kernel of the in-plane geometry, and the depletion length at the etched edges, $150$~nm, is the single geometric parameter. At $B=0$ and $T=0$ this gives the density profile of Chklovskii \emph{et al.}~\cite{Chklovskii1992}, with $n_{\rm el}(0)=1.20\times10^{11}$~cm$^{-2}$, the density of the experiment~\cite{Sirt2025NJP}, an average density of $1.11\times10^{11}$~cm$^{-2}$ and a Fermi energy of $4.3$~meV in the center. At finite $B$ the density, the thermodynamic density of states and the conductivities are evaluated in the self-consistent Born approximation (SCBA) for spin-degenerate Landau levels and Gaussian impurity potentials of range $R=10$~nm~\cite{Guven2003,Siddiki2004}. The single disorder parameter, the SCBA level width, is fixed by the measured mobility: $\Gamma/\hbar\omega_c=0.023$ at $10$~T reproduces $\mu=3.8\times10^6$~cm$^2$/Vs at the mean density; with the $\Gamma\propto\sqrt{B}$ of short-range scatterers this is $\Gamma=0.40\,{\rm meV}\sqrt{B/10\,{\rm T}}$, i.e.\ $0.2$~meV at $2.5$~T. The grid has $1000$ intervals ($10$~nm); the nonlinear equations are solved by Newton iteration while the temperature is lowered in steps from $0.3E_F$ to $k_BT=0.006E_F$ ($0.3$~K). GaAs parameters are used ($m^\ast=0.067m_e$, $\kappa=12.4$). The spin-split plateaus require the exchange-enhanced $g$ factor and are not considered; the arguments below do not depend on $\nu$.

The local conductivity tensor, $\sigH(x)=[\nu(x)-\Delta\sigma_{\rm H}(x)]\,e^2/h$ and $\sigl(x)$, follows from the same SCBA spectral functions~\cite{Guven2003,Siddiki2004}. Nonlocal effects on the scale of the magnetic length are simulated, as in Ref.~\cite{Siddiki2004}, by evaluating the transport coefficients on a tenfold refined grid and averaging them over a length $\lambda=40$~nm. In the translation-invariant bar a current $I$ along $y$ produces a Hall field $E_x(x)=\rho_{\rm H}(x)E_y^0/\rho_\ell(x)$ with a constant $E_y^0$ fixed by $I$~\cite{Guven2003}; the local-equilibrium feedback of G\"uven and Gerhardts~\cite{Guven2003}, in which the position-dependent electrochemical potential $\mus(x)$ re-enters the density, is iterated to self-consistency at each temperature step. In the program the Hall voltage is prescribed and the current follows; we quote the current $I=V_{\rm H}/R_{\rm H}$ with the quantized $R_{\rm H}=h/2e^2$ of the plateau. The Hall and longitudinal resistances follow from $\mus(d)-\mus(-d)$ and $E_y^0$.

Two current densities are distinguished. The \emph{transport} current is
\begin{equation}
j_{\rm tr}(x)=\frac{\sigH(x)}{e}\,\frac{d\mus}{dx},
\label{eq:jtr}
\end{equation}
the current driven by the gradient of the electrochemical potential; it integrates to the terminal current, vanishes in equilibrium, and carries the dissipation~\cite{Guven2003,Gerhardts2008}. The \emph{total drift} current, which is the quantity considered in Ref.~\cite{Sirt2024arXiv},
\begin{equation}
j_{\rm tot}(x)=\frac{\sigH(x)}{e}\,\frac{dV}{dx}\simeq\frac{e}{h}\,\nu(x)\,\frac{dV}{dx},
\label{eq:jtot}
\end{equation}
contains in addition the equilibrium current driven by the gradient of the confinement, which is present at $I=0$ and integrates to zero across the bar; the second form neglects the SCBA correction $\Delta\sigma_{\rm H}$, which vanishes in the strips and stays below $0.013\,e^2/h$ ($2\%$ of $\bar\nu$) in the compressible regions at the fields considered. The two differ by the equilibrium part, $j_{\rm tot}=j_{\rm eq}+j_{\rm tr}$, with $j_{\rm eq}=(e/h)\nu\,d(V-\mus)/dx$. This $j_{\rm eq}$ is the drift part of the equilibrium current, the one carried by the guiding centres in the gradient of the Landau-level energy, and it is what the Kubo Hall conductivity of the SCBA produces. The complete equilibrium current density is the curl of the local magnetization, $j_{\rm eq}^{\rm full}=dM/dx$ with $M(x)=-\partial\Omega/\partial B$ at the local chemical potential $\mu-V(x)$~\cite{Geller1994,Geller1995,Cooper1997}; it contains in addition the density-gradient term $(e\hbar/m^\ast)\sum_n(n+\tfrac12)\,dn_n/dx$ of the compressible regions, which flows opposite to the drift current and does not integrate to zero over an edge region. We evaluate it from the self-consistent profiles with the occupations of the SCBA solution (Sec.~\ref{sec:currents}): the complete equilibrium current of an edge region is the orbital magnetization of the bulk, $M_{\rm bulk}$, as it must be, and is two to five times smaller than the drift part. The transport current and all terminal quantities are unaffected; the total current of an edge region is $M_{\rm bulk}$ plus the transport part. Figures~\ref{fig:profiles} and \ref{fig:eqmaps} show the drift part, Fig.~\ref{fig:profiles}(c,f) also the complete equilibrium current.

\subsection{Two-dimensional six-terminal device}

\begin{figure}
\includegraphics[width=\columnwidth]{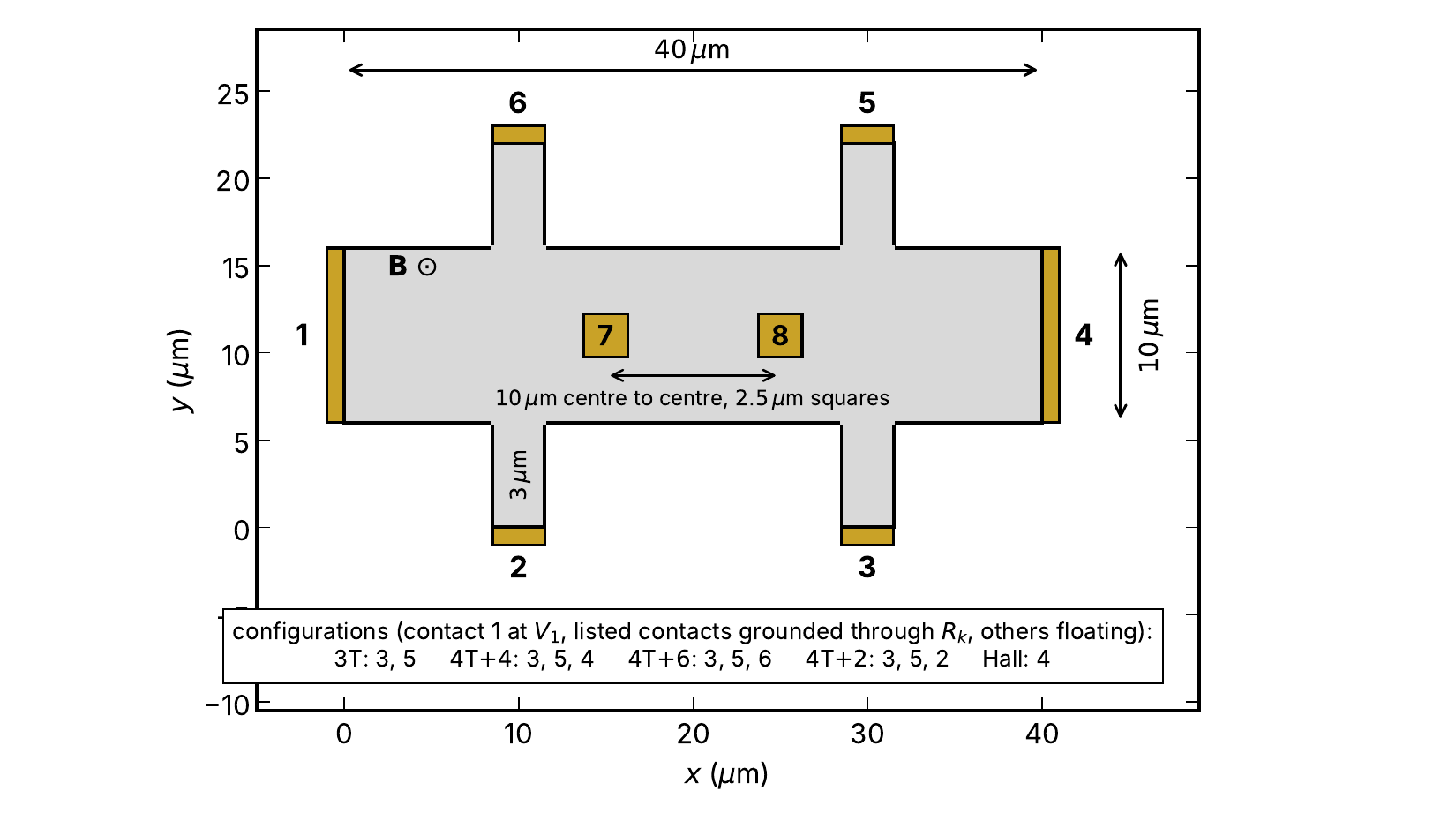}
\caption{The six-terminal device of Ref.~\cite{Sirt2025NJP} as modelled: bar of width $10\um$ and length $40\um$, four arms of width $3\um$ and length $6\um$ at $x=10$ and $30\um$, end contacts $1$ and $4$, arm contacts $2$, $3$ (lower) and $6$, $5$ (upper), and the two interior contacts $7$ and $8$ ($2.5\um$ squares, $10\um$ apart on the axis), all contacts behind the series resistances $R_k$ of Table~I of Ref.~\cite{Sirt2025NJP}. The box lists the measured configurations.}
\label{fig:geometry}
\end{figure}

To compute the current partition we solve the two-dimensional local transport problem $\bm\nabla\cdot(\hat\sigma\bm\nabla\mus)=0$ on the device of Fig.~5 of Ref.~\cite{Sirt2025NJP}: a bar of width $10\um$ and length $40\um$ with $3\um$ wide, $6\um$ long arms at $x=10$ and $30\um$, numbered $1$ (left end), $2$ and $3$ (lower arms), $4$ (right end), $5$ and $6$ (upper arms). The sample also carries two interior contacts, $2.5\um$ squares on the axis of the bar $10\um$ apart, which were used in the interior-contact experiments~\cite{Kendirlik2017,Sirt2025APL} and are floating in the partition measurements; they are included as floating equipotentials (contacts $7$ and $8$ at $x=15$ and $25\um$), and the 2DES is taken to be depleted at their boundary as at an etched edge, so that incompressible strips encircle them whenever the bulk filling factor exceeds the integer, as in the analysis of Ref.~\cite{Sirt2025APL}. We use linear finite elements on a $100$~nm mesh ($4.8\times10^4$ nodes). The averaged filling factor $\bar\nu(x,y)$ and the local conductivities are taken from the translation-invariant SCBA solution as functions of the distance to the nearest etched edge; the strips thus follow the arms and the corners, and the contact interfaces are not treated as edges. Since the strips can be narrower than a mesh cell, each cell is treated as a locally layered medium, with the layers parallel to the nearest edge, and is given the exact homogenized tensor of the layers it covers,
\begin{equation}
\begin{split}
\bar\sigma_{nn}&=\langle\sigl^{-1}\rangle^{-1},\qquad
\bar\sigma_{\rm H}=\frac{\langle\sigH/\sigl\rangle}{\langle\sigl^{-1}\rangle},\\
\bar\sigma_{tt}&=\langle\sigl\rangle+\langle\sigH^2/\sigl\rangle-\frac{\langle\sigH/\sigl\rangle^2}{\langle\sigl^{-1}\rangle},
\end{split}
\label{eq:homog}
\end{equation}
where $n$ and $t$ denote the directions normal and tangential to the layers, $\bar\sigma_{\rm H}$ the antisymmetric part of the cell tensor, and $\langle\cdots\rangle$ the average over the distance interval covered by the cell. Equation~(\ref{eq:homog}) follows from the continuity of the tangential field and of the normal current across the layers. It preserves the leakage resistance across a strip exactly, whatever the position of the strip relative to the mesh; an arithmetic cell average instead lets a narrow strip insulate or leak depending on whether it falls into one cell or straddles two, and modulates the results with the period of the strip crossing one cell, about $0.1$~T. In Eq.~(\ref{eq:homog}) $\sigl$ is bounded below by a floor of $10^{-6}\,e^2/h$, which stands for the residual conductivity of the incompressible regions; the ratios and the floating potentials below change by less than $0.5\%$ and $0.01$~mV, respectively, when the floor is varied between $10^{-7}$ and $10^{-4}\,e^2/h$. With Eq.~(\ref{eq:homog}) a $50$~nm mesh ($1.9\times10^5$ nodes) changes the ratios by less than $0.5\%$ and the floating potentials by less than $0.002$~mV at all fields. The intrinsic six-terminal conductance matrix (all $R_k\to0$) satisfies the Onsager--Casimir relation $G_{kl}(B)=G_{lk}(-B)$ and current conservation to $10^{-11}$ relative to the largest matrix element. The SCBA code, the two-dimensional solver, the sweep data and the scripts that produce every figure and every number quoted in the text are provided as Supplemental Material~\cite{SM}. The device calculation is a linear-response calculation: the conductivities are those of the translation-invariant solution at $I=0$, and the bias-induced displacement of the strips of Sec.~\ref{sec:currents} is not fed back in two dimensions. On the plateau this cannot affect the partition, which by Sec.~\ref{sec:formal} does not depend on the current distribution; between the plateaus and for the floating potentials it is an uncontrolled correction, of the order of the $10\%$ asymmetry that Table~\ref{tab:edges} shows at the $60$~nA of the experiment. Figure~\ref{fig:geometry} shows the device and the measured configurations; note that the translation-invariant bar of Sec.~\ref{sec:1d} and Figs.~\ref{fig:profiles}--\ref{fig:edges} is oriented along $y$ with the cross section along $x$, whereas the device is oriented along $x$ with the cross section along $y$. Each contact pad is an equipotential $U_k$ connected through the series resistance $R_k$ of Table~I of Ref.~\cite{Sirt2025NJP} ($R_1=R_2=R_5=2420$, $R_3=R_6=2400$, $R_4=2350~\Omega$) to the applied voltage $V_1=1$~mV, to ground including the amplifier offsets $V_3=-21$ and $V_5=-18~\mu$V, or left floating. Reference~\cite{Sirt2025NJP} obtained these $R_k$ from two-terminal measurements at $B=0$ and then ``fine-tuned'' them against the field-dependent data with its models. We take them as given and adjust nothing else: the clean solution contains no free parameter. By Sec.~\ref{sec:formal} the plateau partition depends on the $R_k$ and on nothing else in any model with $\sigl=0$, so that the plateau agreement below is a consistency check rather than a test; the tests are the regime between the plateaus and the floating-contact potentials of Sec.~\ref{sec:probes}. The four measured configurations are reproduced: contacts $3$ and $5$ grounded (3T), and additionally $4$, $6$ or $2$ grounded (4T$+4$, 4T$+6$, 4T$+2$). The homogeneous model contains neither the long-range density fluctuations of the real sample nor the spin splitting. To show how the measured plateau emerges from the clean solution we also present, as a second variant, a minimal phenomenology for both, applied to the self-consistent profiles before they enter the transport problem: (i)~regions with $|\bar\nu-n|<\delta$ ($n=2,4$) are treated as quantized, $\sigl=0$ and $\sigH=ne^2/h$, which is the localization of the compressible puddles of a long-range disorder potential; (ii)~$\sigl$ is averaged over $\lambda_{\rm eff}=300$~nm along the normal to the edge, so that strips narrower than the lateral displacement imposed on them by the same disorder do not insulate; (iii)~$\sigl$ of the remaining compressible regions is multiplied by $\alpha=0.25$, which accounts for the spin-split minima that reduce the measured dissipation between the even plateaus. We use $\delta=0.2$, the value needed for the interior of the $3\um$ arms, whose filling factor is $4\%$ below that of the bar center, to be quantized up to the measured end of the plateau; $\delta$ is thus selected with the measured plateau end, and the end moves with it ($2.54$, $2.62$ and $2.70$~T for $\delta=0.15$, $0.2$ and $0.25$), whereas the suppression between the plateaus does not ($0.83$--$0.94$ at $3.0$~T for all three). All three parameters correspond to density fluctuations of the order of $10\%$. The clean solution remains the primary result; the phenomenology has no influence on the arguments of Secs.~\ref{sec:formal} and \ref{sec:currents}. The sign of $B$ is chosen such that the currents circulate $1\to2\to3\to\cdots$, as for $B>0$ in Ref.~\cite{Sirt2025NJP}; reversing $B$ transposes $\hat\sigma$ and interchanges the roles of contacts $3$ and $5$. Kirchhoff's law is satisfied to $10^{-12}$, and at $B=0$ with a scalar conductivity the symmetric 3T configuration gives $-I_3/I_1=0.504$, $-I_5/I_1=0.496$, the small asymmetry coming from $R_3\neq R_5$ and the offsets.

\section{Results}
\label{sec:results}

\subsection{Plateau and strips}

\begin{figure*}
\includegraphics[width=\textwidth]{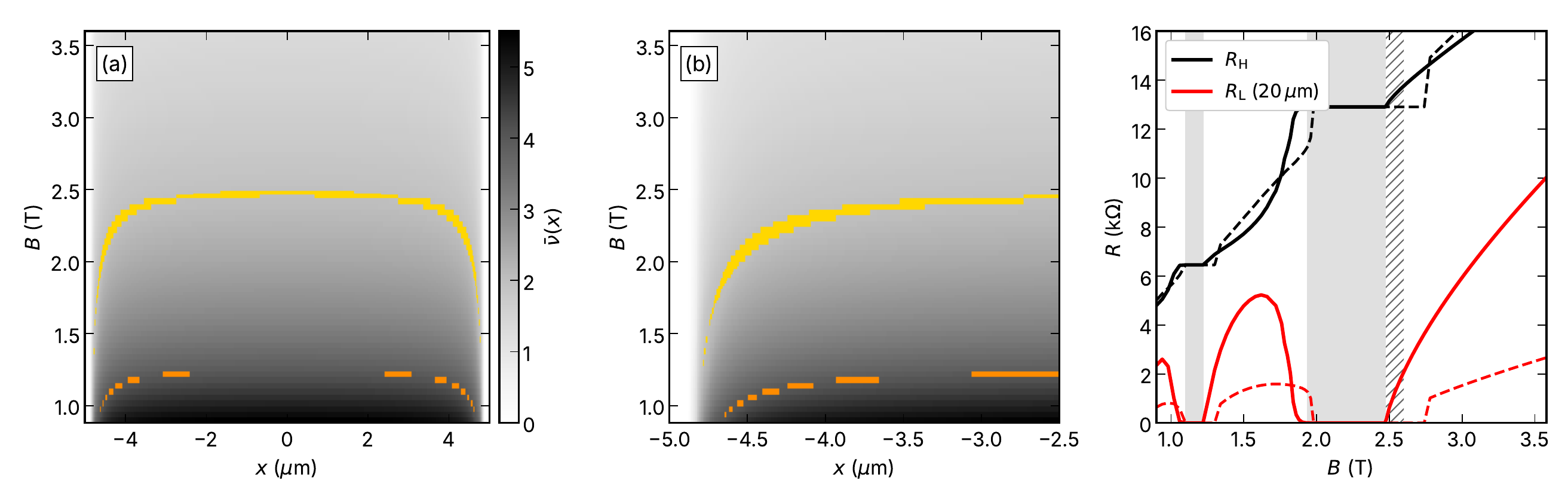}
\caption{(a)~Averaged local filling factor $\bar\nu(x)$ across the $10\um$ bar as a function of $B$ (gray scale); incompressible regions with $|\bar\nu-2|<0.01$ (yellow) and $|\bar\nu-4|<0.01$ (orange). (b)~Enlargement of the left edge. (c)~Hall resistance $R_{\rm H}$ (black) and longitudinal resistance $R_{\rm L}$ (red, for the $20\um$ probe separation) from the translation-invariant local-Ohm solution with the clean SCBA conductivities (solid) and with the disorder phenomenology of Sec.~\ref{sec:model} (dashed). Shaded intervals: calculated plateaus of the clean solution, $R_{\rm L}<10^{-4}h/e^2$; hatched: the measured $\nu=2$ plateau, $2.3$--$2.6$~T~\cite{Sirt2025NJP}.}
\label{fig:sweep}
\end{figure*}

Figure~\ref{fig:sweep} shows the evolution of the incompressible strips and the resistances. On the low-field side of the $\nu=2$ plateau two ICSs sit near the edges ($300$~nm wide, $0.8\um$ from the edge at $2.3$~T); with increasing $B$ they move inward and widen, and at $B\approx2.47$~T, where the center filling factor reaches $2$, they merge into a bulk ICS $4\um$ wide. The plateau, defined by $R_{\rm L}<10^{-4}h/e^2$ for the $20\um$ probe separation, extends from $1.94$ to $2.46$~T, and the $\nu=4$ plateau from $1.10$ to $1.22$~T; between the plateaus $R_{\rm L}$ reaches $5$~k$\Omega$ and $R_{\rm H}$ interpolates smoothly between the quantized values. In the experiment~\cite{Sirt2025NJP} the $\nu=2$ plateau lies at $2.3$--$2.6$~T and the maxima of $R_{\rm L}$ between the plateaus are $1$--$2$~k$\Omega$; the clean calculation thus places the plateau $0.3$~T too low and overestimates the dissipation between the plateaus by a factor of $3$--$6$. With the disorder phenomenology (dashed lines) the translation-invariant plateau becomes $2.02$--$2.74$~T and $R_{\rm L}$ between the plateaus $1$--$2$~k$\Omega$; the measured plateau width is recovered only in the two-dimensional device, where the arms and the interior contacts end it (Secs.~\ref{sec:partition} and \ref{sec:probes}). The upper end of the calculated plateau coincides with the field at which the clean bulk becomes compressible; its extension to higher $B$ in real samples is due to long-range disorder, which breaks the compressible bulk into puddles separated by a percolating incompressible network~\cite{Huckestein1995,Ilani2004,Siddiki2004}, and is not contained in the homogeneous model. The low-field onset lies where the edge strips become wider than the averaging length $\lambda$; it is sensitive to $\Gamma$ and, numerically, to the grid, which must resolve the strips (with $20$~nm instead of $10$~nm cells the resistances acquire spurious steps between $1.7$ and $1.9$~T). These offsets do not affect the arguments below, which concern the regimes (edge strips, bulk strip, compressible bulk) rather than the fields at which they occur.

\subsection{Equilibrium current and transport current}
\label{sec:currents}

\begin{figure*}
\includegraphics[width=0.9\textwidth]{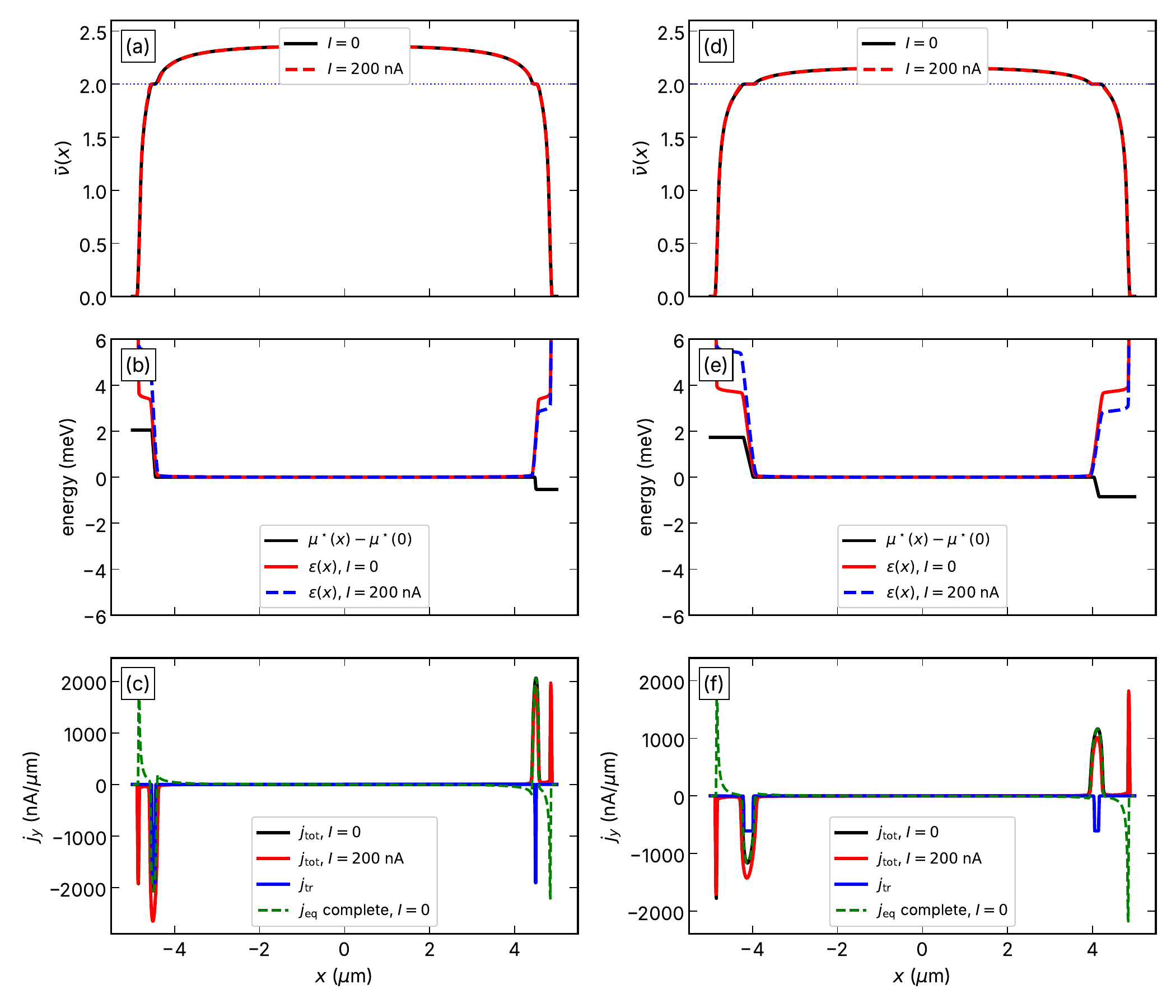}
\caption{Translation-invariant bar at $B=2.3$~T (left column, two edge strips) and $B=2.46$~T (right column, inner strips $1.3$--$2.6\um$ from the center, just before they merge). (a,d)~Averaged filling factor in equilibrium (black) and with $I=200$~nA (red dashed). (b,e)~Electrochemical potential $\mus(x)-\mu_0$ with $I=200$~nA (black), and the Landau-level energy $\varepsilon(x)=V(x)-V(0)$ in equilibrium (red) and with current (blue dashed). (c,f)~Total drift current density $j_{\rm tot}$ in equilibrium (black) and with current (red), the transport current density $j_{\rm tr}$ (blue) for $I=200$~nA, and the complete equilibrium current density $dM/dx$ (green dashed), which adds to the drift part the counter-flowing density-gradient current of the compressible regions~\cite{Geller1994} and integrates over each edge region to the bulk magnetization. The total current is opposite at the two edges; the transport current has the same sign in both strips.}
\label{fig:profiles}
\end{figure*}

Figure~\ref{fig:profiles} compares the two current densities of Eqs.~(\ref{eq:jtr}) and (\ref{eq:jtot}) for the two characteristic situations of the plateau. In equilibrium ($I=0$) the total drift current density is confined to the regions where $\nu\,dV/dx\neq0$, i.e.\ to the incompressible strips and the depletion zones, it is opposite at the two edges, and it integrates to zero across the bar. Over one edge region it integrates to $331$--$362$~nA, close to $(e/h)\,\nu\,\hbar\omega_c$, the current carried by $\nu$ filled Landau levels across the potential drop $\hbar\omega_c$ of one strip. The complete equilibrium current [green dashed line in Fig.~\ref{fig:profiles}(c,f)] adds the density-gradient current of the compressible regions, which flows in the opposite sense in the depletion zone and on the bulk side of the strip~\cite{Geller1994}, and reduces the equilibrium current of an edge region to the orbital magnetization of the bulk: $M_{\rm bulk}=60$, $110$ and $147$~nA at $2.10$, $2.30$ and $2.46$~T, which follows the de Haas--van Alphen sawtooth of the bulk filling and is largest just at integer filling. This equilibrium current is chiral in the sense of Ref.~\cite{Sirt2024arXiv}, and it would be chiral in any model in which the Landau-level energy rises towards both edges; it is also unobservable by terminal measurements, because no cross section of the bar carries a net equilibrium current.

The transport current is different. With $I\neq0$ the electrochemical potential $\mus(x)$ drops monotonically from one edge to the other, across both edge strips at $2.3$~T [Fig.~\ref{fig:profiles}(b)] and across the two inner strips at $2.46$~T [Fig.~\ref{fig:profiles}(e)]. Since $\sigH>0$ everywhere, $j_{\rm tr}=\sigH\,d\mus/dx/e$ has the same sign wherever $\mus$ drops: the transport current flows in the same direction in both edge strips [Fig.~\ref{fig:profiles}(c)] and in both inner strips [Fig.~\ref{fig:profiles}(f)]; at $2.47$~T, where the inner strips merge, it flows across the single bulk strip (Sec.~\ref{sec:partition}). It is unidirectional. The total current with $I\neq0$ is the sum of the two, so that at $200$~nA the current in the strip on the high-$\mus$ side grows while the oppositely directed current in the strip on the low-$\mus$ side shrinks [Fig.~\ref{fig:profiles}(c)]; the total current of the low-$\mus$ edge remains opposite to that of the high-$\mus$ edge as long as the transport current of the low side, $(1-f_{\rm high})I$, with the share $f_{\rm high}$ of the high-$\mus$ strip listed in Table~\ref{tab:edges}, is smaller than $M_{\rm bulk}$. This holds over the whole computed range $I\le200$~nA; with $f_{\rm high}$ frozen at its $200$~nA value the crossover would lie at $I\approx270$, $340$ and $390$~nA at $2.10$, $2.30$ and $2.46$~T, and since $f_{\rm high}$ keeps growing with $I$ these are lower bounds, in any case well above the $66$~nA of the partition experiment. The integrated total current equals $I$ to better than $1\%$.

\begin{table}
\caption{Translation-invariant bar: current in the two edge regions (integrated total drift current $j_{\rm tot}$ over each half of the bar, nA) at $I=0$ and at three imposed currents $I$, for three fields in the $\nu=2$ plateau. The share is $I_{\rm tr,high}/(I_{\rm tr,high}+I_{\rm tr,low})$, the fraction of the integrated transport current $j_{\rm tr}$ carried by the half of the bar on the high-$\mus$ side; $\Delta\mus_{\rm high}$ and $\Delta\mus_{\rm low}$ are the drops of the electrochemical potential over the two halves (meV), essentially across the strip on each side. The share differs slightly from $\Delta\mus_{\rm high}/(\Delta\mus_{\rm high}+\Delta\mus_{\rm low})$ because a small part of each drop occurs where $\sigH\neq2e^2/h$. The currents listed are drift currents; the complete equilibrium current of an edge region, including the density-gradient part of the compressible regions, is $M_{\rm bulk}=60$, $110$ and $147$~nA at the three fields, independent of $I$ to $0.1$~nA, and the complete total current of an edge is $M_{\rm bulk}$ plus (high side) or minus (low side) the transport current of that side.}
\label{tab:edges}
\begin{ruledtabular}
\begin{tabular}{ccccccc}
$B$ & $I$ & $|I_{\rm high}|$ & $|I_{\rm low}|$ & share & $\Delta\mus_{\rm high}$ & $\Delta\mus_{\rm low}$\\
(T) & (nA) & (nA) & (nA) & & (meV) & (meV)\\
\hline
2.10 & 0   & 331 & 331 & --   & 0 & 0\\
2.10 & 10  & 336 & 326 & 0.55 & 0.067 & 0.062\\
2.10 & 60  & 368 & 309 & 0.62 & 0.487 & 0.287\\
2.10 & 200  & 488 & 290 & 0.78 & 2.047 & 0.534\\
2.30 & 0   & 350 & 350 & --   & 0 & 0\\
2.30 & 10  & 355 & 345 & 0.51 & 0.066 & 0.063\\
2.30 & 60  & 383 & 323 & 0.56 & 0.432 & 0.342\\
2.30 & 200  & 483 & 284 & 0.68 & 1.733 & 0.849\\
2.46 & 0   & 362 & 362 & --   & 0 & 0\\
2.46 & 10  & 367 & 357 & 0.51 & 0.065 & 0.064\\
2.46 & 60  & 394 & 334 & 0.54 & 0.417 & 0.357\\
2.46 & 200  & 486 & 286 & 0.62 & 1.603 & 0.979\\
\end{tabular}
\end{ruledtabular}
\end{table}

\begin{figure*}
\includegraphics[width=0.9\textwidth]{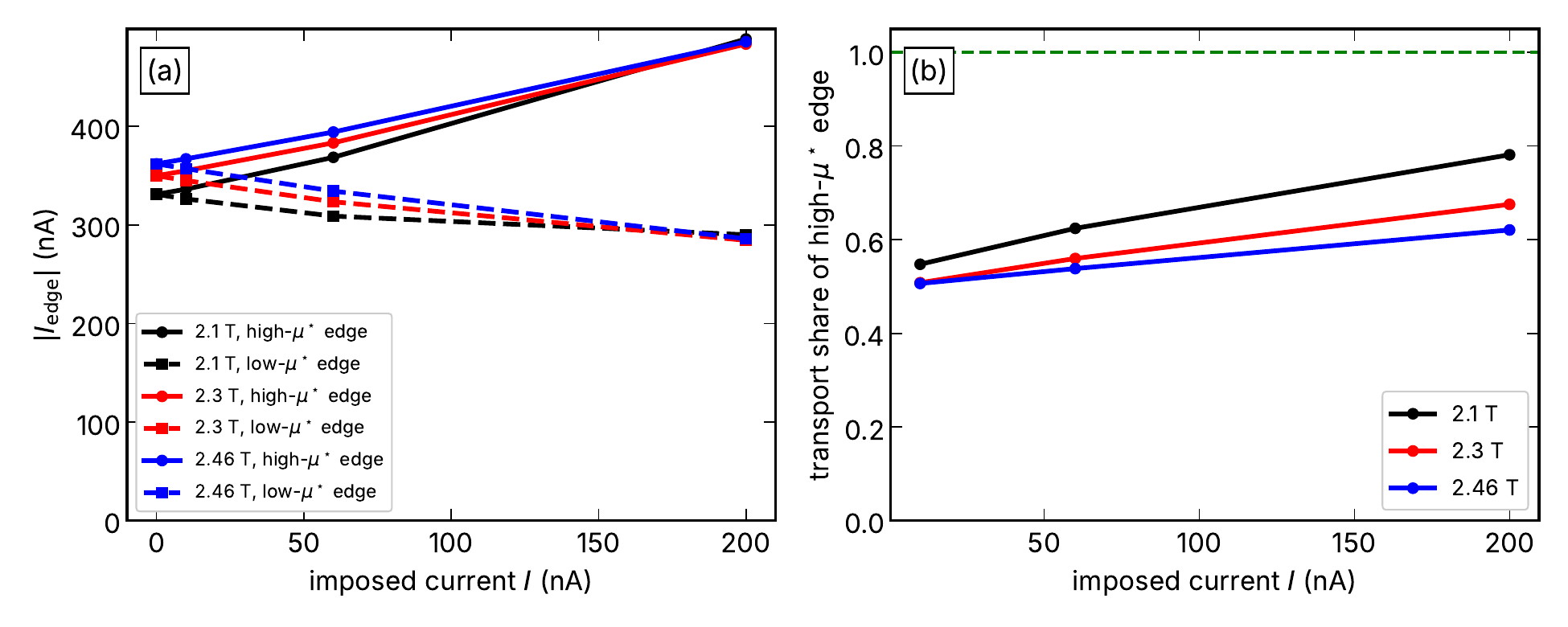}
\caption{(a)~Integrated total drift current in the edge region on the high-$\mus$ side (solid) and on the low-$\mus$ side (dashed) as a function of the imposed current, at $2.1$ (black), $2.3$ (red) and $2.46$~T (blue). (b)~Share of the transport current carried by the high-$\mus$ strip; the green line marks the ``one-edge'' limit of Ref.~\cite{Sirt2024arXiv}. In linear response the two strips share the current equally; the asymmetry grows with $I$ and is strongest in the edge-strip regime.}
\label{fig:edges}
\end{figure*}

Table~\ref{tab:edges} and Fig.~\ref{fig:edges} quantify how the current induced by the bias is shared between the two edges. In linear response the two strips carry the transport current equally, by the mirror symmetry of the bar, and the Hall voltage drops in equal parts across them; at $I=10$~nA the share of the high-$\mus$ strip is already $0.51$--$0.55$, the $I\to0$ limit being $0.50$. With increasing $I$ the self-consistent feedback widens the strip on the high-$\mus$ side and narrows the other; the share grows to $0.62$--$0.78$ at $200$~nA and the asymmetry is strongest in the edge-strip regime. This is the self-consistent asymmetry of the strips under bias that underlies the current-direction and bias dependence discussed in Refs.~\cite{Siddiki2009,Siddiki2010}, and it is what SFM observes: Panos \emph{et al.}~\cite{Panos2014} found on the low-field side of the $\nu=2$ plateau two Hall-potential drops near the two edges, symmetric at low bias and increasingly asymmetric with bias, and on the high-field side a drop spread across the bulk; Dahlem \emph{et al.}~\cite{Dahlem2010} report that near integer filling the Hall potential drops ``on both sides of the Hall bar almost symmetrically at the expected positions of the innermost incompressible strips'', and that one-sided drops occur when the depletion at a contact interface decouples one edge from that contact, a property of the contact rather than of the bulk current distribution.

We can now place the decomposition of Ref.~\cite{Sirt2024arXiv}. There the current density is split into a persistent part, carried by states below a reference chemical potential $\mu_0$, and an imposed part, carried by states between $\mu_0$ and the higher contact potential, and $\mu_0$ is chosen as the chemical potential of the low-potential edge. With this choice every state on the low-potential edge is by definition persistent, and the imposed current is by definition confined to the high-potential edge; the reduction of the total current on the low-potential side is booked as a reduction of the persistent current. Choosing instead the high-potential edge as reference puts the whole imposed current on the low-potential edge, as a deficit of the oppositely directed current there (the hole representation of Ref.~\cite{Komiyama1996}); choosing the mean puts half on each. The decomposition is a representation, as discussed in general by Komiyama and Hirai~\cite{Komiyama1996}, and no measurement distinguishes its variants, because the total current density is the same in all of them. The physically meaningful question is how the current density changes when the bias is switched on, $j_{\rm tot}(x;I)-j_{\rm tot}(x;0)$, and Table~\ref{tab:edges} answers it: at $2.46$~T and $200$~nA the change is $+124$~nA in the high-side strip and $+76$~nA in the low-side strip, both in the direction of the net current. Integrated over either edge region this is $j_{\rm tr}$, and it is unidirectional: the drift part of the equilibrium current, $(e/h)\nu\,d(V-\mus)/dx$ integrated over either edge, changes by less than $1$~nA ($0.3\%$) between $I=0$ and $200$~nA at all three fields, because the self-consistent $V(x)$ and $\mus(x)$ shift together, and the complete equilibrium current of an edge region, $M_{\rm bulk}$, changes by less than $0.1$~nA, because the bulk occupation does not. Pointwise the statement is weaker: the strips move and widen under bias [Fig.~\ref{fig:profiles}(c)], so that $j_{\rm eq}(x;I)-j_{\rm eq}(x;0)$ is a displacement profile of alternating sign, as large locally as $j_{\rm tr}$, whose integral over each edge region vanishes. The bias-induced change of the current carried by each edge is $j_{\rm tr}$; the bias-induced change of the current density at a given point is $j_{\rm tr}$ plus this displacement. The total edge currents of Table~\ref{tab:edges} therefore have exactly the form of Ref.~\cite{Sirt2024arXiv}, $I_{\rm high}=I_{\rm pers}+I$ and $I_{\rm low}=I_{\rm pers}$ with an $I_{\rm pers}$ that decreases with $I$: at $2.3$~T, with the complete equilibrium current, from $110$~nA at $I=0$ to $45$~nA at $200$~nA (with the drift part alone from $350$ to $284$~nA). Statements (iv) and (v) of Ref.~\cite{Sirt2024arXiv} thus hold for the total currents of the self-consistent solution as well. What the solution adds is their meaning: the decrease of $I_{\rm pers}$ is the transport current of the low-$\mus$ strip, $65$~nA at $200$~nA, flowing against an equilibrium current that itself does not change, and the current booked on the high-$\mus$ edge, $I$, is the transport current of that strip, $135$~nA, plus the $65$~nA that the other strip carries in the same direction. The conclusion of Ref.~\cite{Sirt2024arXiv} that this contradicts G\"uven and Gerhardts~\cite{Guven2003} therefore rests on the choice of reference: what is presented there as a finding in conflict with Ref.~\cite{Guven2003} is a convention, and the same convention applied to the solution of Ref.~\cite{Guven2003} produces the same statement. Reference~\cite{Sirt2024arXiv} also describes the local Ohm's law of Ref.~\cite{Guven2003}, $j\propto d\mu/dy$, as an assumption. It is the linear response of the electron system, and it is the only ingredient of the chiral model as well: Eq.~(3) of Ref.~\cite{Sirt2025NJP} is $\sigH\bm\nabla\mus/e$ integrated across the bar. The two papers use the same relation, Ref.~\cite{Sirt2025NJP} integrated across the bar and Ref.~\cite{Guven2003} locally; the integrated form fixes the terminal current and, by Sec.~\ref{sec:formal}, nothing about its distribution. The self-consistent solution confirms Ref.~\cite{Guven2003}. The dependence on the reference can be written down. With $\nu=2$ in both strips, the states of the high-$\mus$ strip whose energies lie between $\mu_0$ and $\mu_S$ carry $(2e/h)(\mu_S-\mu_0)$, and the states missing between $\mu_D$ and $\mu_0$ in the low-$\mus$ strip contribute $(2e/h)(\mu_0-\mu_D)$ in the same direction, so that the fraction of the imposed current booked on the high-$\mus$ edge is
\begin{equation}
f_{\rm high}=\frac{\mu_S-\mu_0}{\mu_S-\mu_D},
\label{eq:fhigh}
\end{equation}
equal to $1$ for $\mu_0=\mu_D$ (Ref.~\cite{Sirt2024arXiv}), to $0$ for $\mu_0=\mu_S$, and to $\Delta\mus_{\rm high}/(\Delta\mus_{\rm high}+\Delta\mus_{\rm low})$, the share of Table~\ref{tab:edges} up to the constant-$\sigH$ approximation noted there, when $\mu_0$ is the electrochemical potential of the compressible bulk between the strips, the natural choice in that it singles out no contact and reproduces the share of the self-consistent solution.

\subsection{Multiterminal current partition}
\label{sec:partition}

\begin{figure*}
\includegraphics[width=\textwidth]{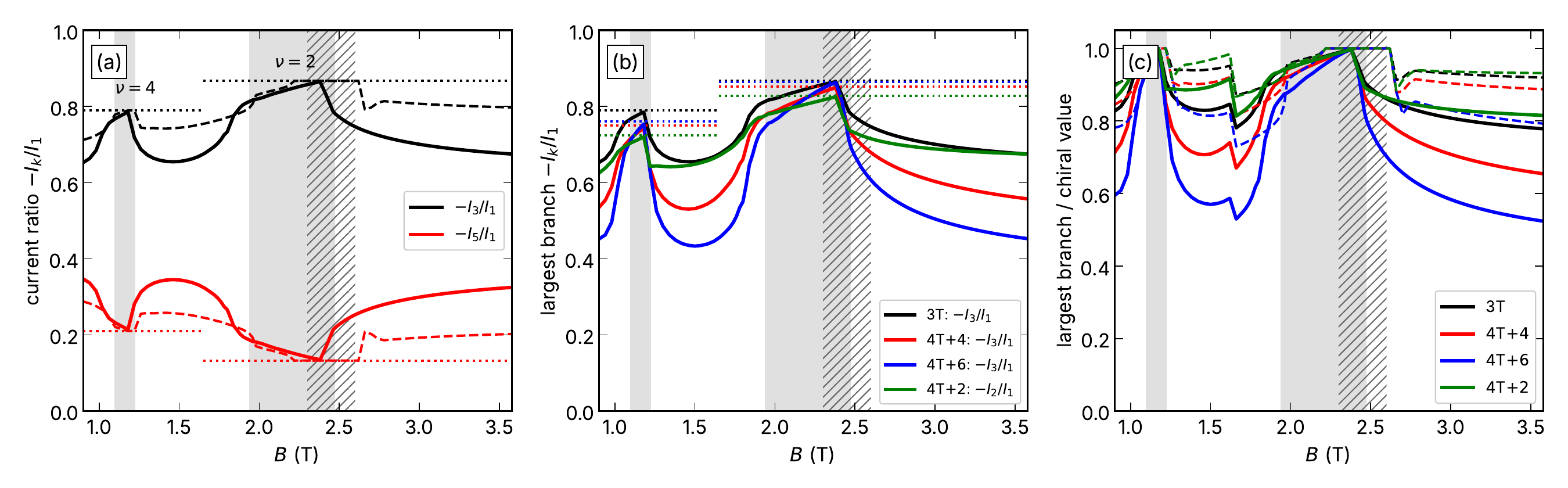}
\caption{Current partition from the two-dimensional local-Ohm solution for the six-terminal device of Ref.~\cite{Sirt2025NJP}. (a)~Three-terminal configuration (contacts 3 and 5 grounded): $-I_3/I_1$ (black) and $-I_5/I_1$ (red) versus $B$ with the clean SCBA conductivities (solid) and with the disorder phenomenology of Sec.~\ref{sec:model} (dashed); dotted lines are the values of the chiral conductance matrix, Eq.~(\ref{eq:LB}) with the measured contact resistances, for $\nu=4$ and $\nu=2$. (b)~Largest-branch ratio for the four configurations, 3T (black), 4T$+4$ (red), 4T$+6$ (blue) and 4T$+2$ (green), clean solution; dotted lines as in (a). (c)~The ratios of (b) divided by their chiral values at the nearest plateau ($\nu=4$ below $1.65$~T, $\nu=2$ above; the step at $1.65$~T is this change of reference), for the clean solution (solid) and with the disorder phenomenology (dashed). Shaded intervals: calculated plateaus of the clean translation-invariant solution; hatched: the measured $\nu=2$ plateau.}
\label{fig:ratios}
\end{figure*}

Figure~\ref{fig:ratios} shows the current partition obtained from the two-dimensional local-Ohm solution with the self-consistent strips, between $0.9$ and $3.6$~T. On the $\nu=2$ plateau the ratios of all four configurations collapse onto the values of the chiral conductance matrix, Eq.~(\ref{eq:LB}), computed with the measured contact resistances: at $2.3$~T the 3T configuration gives $-I_3/I_1=0.857$ and $-I_5/I_1=0.143$ against $0.867$ and $0.133$, and the 4T$+4$, 4T$+6$ and 4T$+2$ configurations give $0.839$, $0.848$ and $0.817$ against $0.852$, $0.865$ and $0.828$; between $2.2$ and $2.4$~T all four lie within $1$--$5\%$ of their chiral values, and at $2.38$~T within $0.3\%$. In the lower part of the plateau, $1.94$--$2.15$~T, where the edge strips are only $50$--$150$~nm wide, the partition is $3$--$14\%$ below the chiral values ($4$--$6\%$ for the 3T configuration), the 4T$+6$ configuration being the most sensitive: the narrow strips are not yet perfect insulators on the scale of the device. The floating interior contacts change the plateau values by about $1\%$ (without them the 3T value at $2.3$~T is $0.849$). The remaining difference comes from the $3\um$ wide arms, whose interiors are compressible in the calculation and add a weakly dissipative path in series with the contacts; in the experiment such a contribution is absorbed in the fitted $R_k$. The collapse is what Sec.~\ref{sec:formal} predicts: once $\sigl=0$ in a region that connects all contacts, the partition is fixed by Eq.~(\ref{eq:LB}) regardless of whether the current flows in two edge strips or in two inner strips. On the narrow $\nu=4$ plateau the same collapse occurs onto the $\nu=4$ values ($0.790$ and $0.210$ for the 3T configuration), to within $1\%$ at $1.18$~T. Reversing the field, i.e.\ transposing $\hat\sigma$, interchanges the roles of the two edges and of contacts $3$ and $5$: at $2.3$~T the clean solution gives $-I_3/I_1=0.149$ and $-I_5/I_1=0.851$, and the solution with the localized bulk $0.139$ and $0.861$, the values of Eq.~(\ref{eq:LB}) for the reversed cyclic order, as in Fig.~6 of Ref.~\cite{Sirt2025NJP}. The current density at the cut between the arms is positive across the full width for both field directions; what reverses is the free edge that carries the potential of contact $1$ ($V_2=V_1'$ for one direction, $V_6=V_1'$ for the other), not the direction of the current anywhere in the bar.

Above $2.4$~T the calculated partition falls below the chiral values before the clean bulk becomes compressible at $2.48$~T, to $0.91$ of the chiral value for the 3T configuration at $2.46$~T and to $0.83$ for the 4T$+6$ configuration, for which the interior contacts open an additional dissipative path once the bulk strip is interrupted by their rims. The reason is geometric: the inner strips have retreated more than $1.5\um$ from the edges and no longer run into the $3\um$ wide arms, whose density in the distance-mapped model is below $\nu=2$, so that the current must cross the compressible arm interiors to reach contacts 3 and 5. In the experiment the partition stays chiral up to $2.6$~T; this is the same shortcoming as the early end of the one-dimensional plateau in Fig.~\ref{fig:sweep}(c), namely the absence in the homogeneous model of the disorder-induced localization that keeps $\sigl=0$ in regions of slightly non-integer average filling~\cite{Huckestein1995,Ilani2004,Siddiki2004}. With the disorder phenomenology of Sec.~\ref{sec:model} (dashed lines in Fig.~\ref{fig:ratios}) the partition of all four configurations is chiral to better than $0.5\%$ from $2.22$ to $2.62$~T, an interval of $0.40$~T that covers the measured $2.3$--$2.6$~T ($0.30$~T), and on the $\nu=4$ plateau, to better than $1.5\%$, from $1.10$ to $1.22$~T. The upper end is now set by the arms and the interior contacts, whose interiors leave the localization window before the bar does: the plateau width seen in a multiterminal measurement is a property of the device, not only of the bulk, and it is narrower than the $2.02$--$2.74$~T of the translation-invariant bar with the same phenomenology.

Between the plateaus the partition is suppressed below the chiral values and depends on the configuration. At $2.7$~T, where the bulk is compressible with $\bar\nu\approx1.8$, the largest branch is reduced to $0.84$ (3T), $0.77$ (4T$+4$), $0.66$ (4T$+6$) and $0.85$ (4T$+2$) of its chiral value, and between the $\nu=4$ and $\nu=2$ plateaus (at $1.6$~T) to $0.84$, $0.73$, $0.58$ and $0.91$: the suppression is largest when the additional grounded contact is closest to the emitting contact 1 ($+6$), then $+4$, then $+2$, which is barely distinguishable from the 3T configuration. This is the ordering reported in Fig.~10 of Ref.~\cite{Sirt2025NJP} and summarized there as ``the current suppression into the contact with the largest current is the greater the closer another grounded contact is to the biased contact''. The magnitude of the suppression is larger in the calculation than in the experiment, where the largest-branch ratios between the plateaus stay within about $5$--$15\%$ of the neighboring plateau values, because the spin-degenerate SCBA model overestimates $\sigl$ in the compressible bulk of this sample, in which the spin-split minima lower the dissipation between the even plateaus: the calculated longitudinal resistance for the $20\um$ probe separation is $4.0$~k$\Omega$ at $2.8$~T and $7.3$~k$\Omega$ at $3.2$~T, against measured values of $2$ and $1.2$~k$\Omega$~\cite{Sirt2025NJP}. Reference~\cite{Sirt2025NJP} describes the same regime by fitting its Drude model with a field-dependent mobility $\mu(B)$ that between the plateaus falls two orders of magnitude below the zero-field value. By their Eq.~(7) this $\mu(B)$ is $\sigH/(\sigl B)$ of a homogeneous conductor, i.e.\ $(R_{\rm H}/R_{xx})(L/W)/B$, which is also how Ref.~\cite{Sirt2025NJP} extracts it from its four-terminal data (its Appendix~C): the measured $R_{xx}$ gives $5$ and $9$~m$^2$/Vs at $2.75$ and $3.2$~T, the clean SCBA bulk $1.4$--$3.3$~m$^2$/Vs and the bulk with $\alpha=0.25$ $6$--$13$~m$^2$/Vs between $2.7$ and $3.4$~T. It is not the carrier mobility: within the SCBA, $\sigl/\sigH$ of a partially filled Landau level is set by the density of states at the Fermi level, not by a scattering time, which is why a fit in terms of $\mu B$ needs a $\mu$ that varies by two orders of magnitude across the plateau. A mobility that must collapse within one tesla and recover between the plateaus is not a measurement of the mobility; it signals that the Drude model is being used outside its range of validity. The ``transition from ballistic transport on the plateaus to diffusive transport between the plateaus'' that Ref.~\cite{Sirt2025NJP} reads from this fit is the transition from $\sigl=0$ to $\sigl\neq0$, which the local-Ohm solution contains without either word; nothing in the agreement depends on the plateau transport being ballistic. With the disorder phenomenology, whose factor $\alpha=0.25$ brings the calculated $R_{\rm L}$ into the measured range, the suppression is $0.79$--$0.94$ of the chiral values [dashed lines in Fig.~\ref{fig:ratios}(c)], in the range of the experiment, with the same ordering: at $3.0$~T the 4T$+6$, 4T$+4$, 4T$+2$ and 3T configurations give $0.83$, $0.91$, $0.94$ and $0.93$. The interior contacts, metallic islands in the compressible bulk, shift the dissipative paths towards the axis of the bar and thereby increase the suppression of the 4T$+6$ configuration and decrease that of 4T$+2$ by $2$--$5\%$; they do not change the ordering. The ordering is thus a property of the geometry and the magnitude a property of $\sigl$; neither requires a model of the current flow beyond the local conductivity tensor, and the local-Ohm solution contains both without an adjustable mobility.

\begin{figure*}
\includegraphics[width=0.6\textwidth]{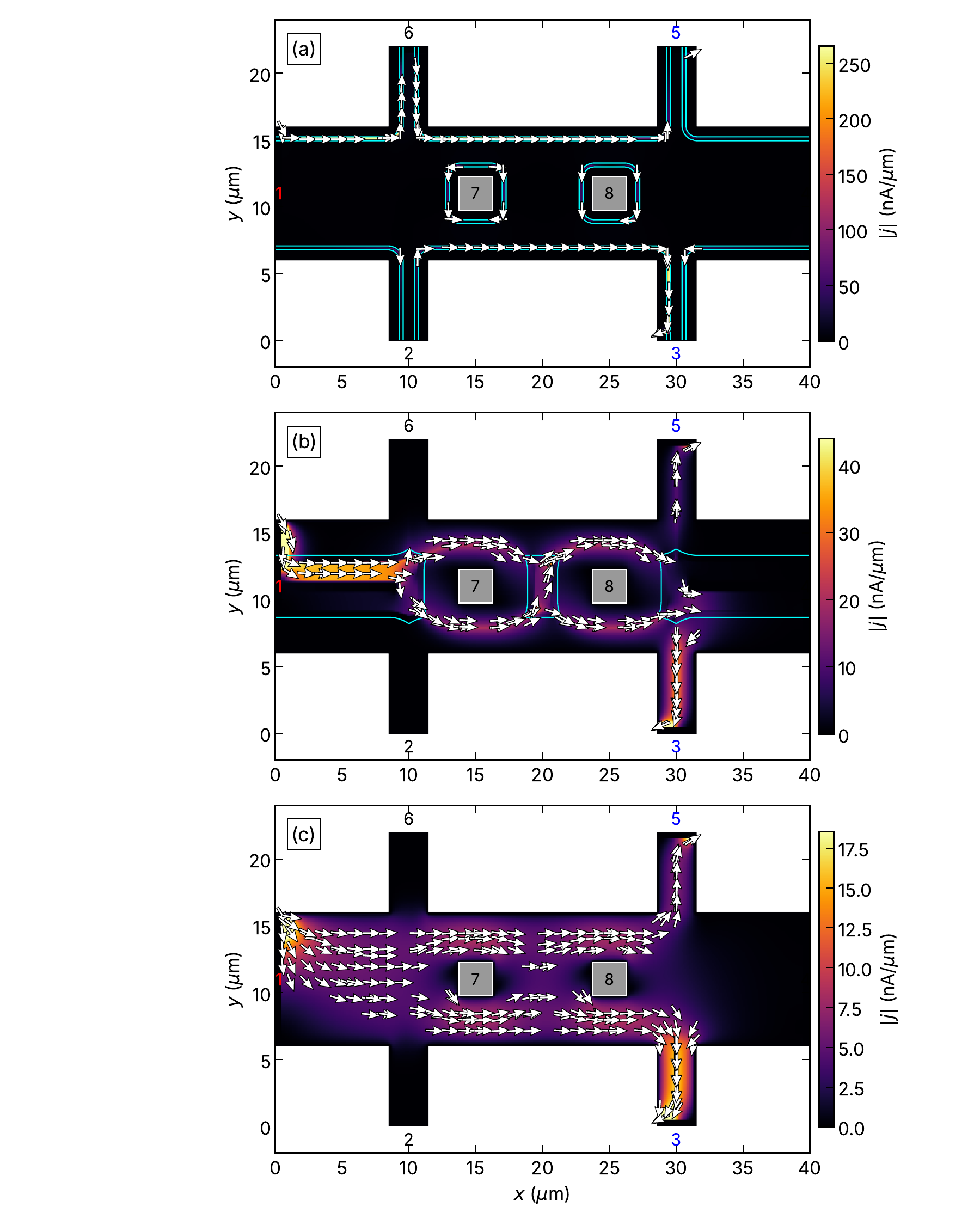}
\caption{Transport current density in the six-terminal device for the 3T configuration (contact 1 at $1$~mV, contacts 3 and 5 grounded through $R_k$, contacts 2, 4, 6 and the interior contacts 7, 8 floating) at (a)~$2.3$~T (two edge strips), (b)~$2.47$~T (bulk strip) and (c)~$2.7$~T (compressible bulk). Colour: $|\bm j|$; arrows: direction of $\bm j$ in the regions that carry the current; cyan lines: $\bar\nu=2$; gray squares: the interior contacts. In (a) the current emitted by contact 1 flows in the same direction along both edge strips, circulates around the floating arm 2, enters contact 3 through the lower strip and contact 5 through the upper strip; beyond the arms, around the floating end 4, only a loop of $3$~nA circulates, carrying no net current, and the interior contacts are isolated by closed incompressible rings and carry no current. In (b) the bulk strip is interrupted by the depleted rims of the interior contacts and the current flows around them through the compressible rims attached to the floating metal, entering the arms through their compressible interiors; in (c) it spreads over the bulk, divides around the interior contacts, and shows the hot spots at the diagonally opposite contact corners.}
\label{fig:maps}
\end{figure*}

\begin{figure*}
\includegraphics[width=0.9\textwidth]{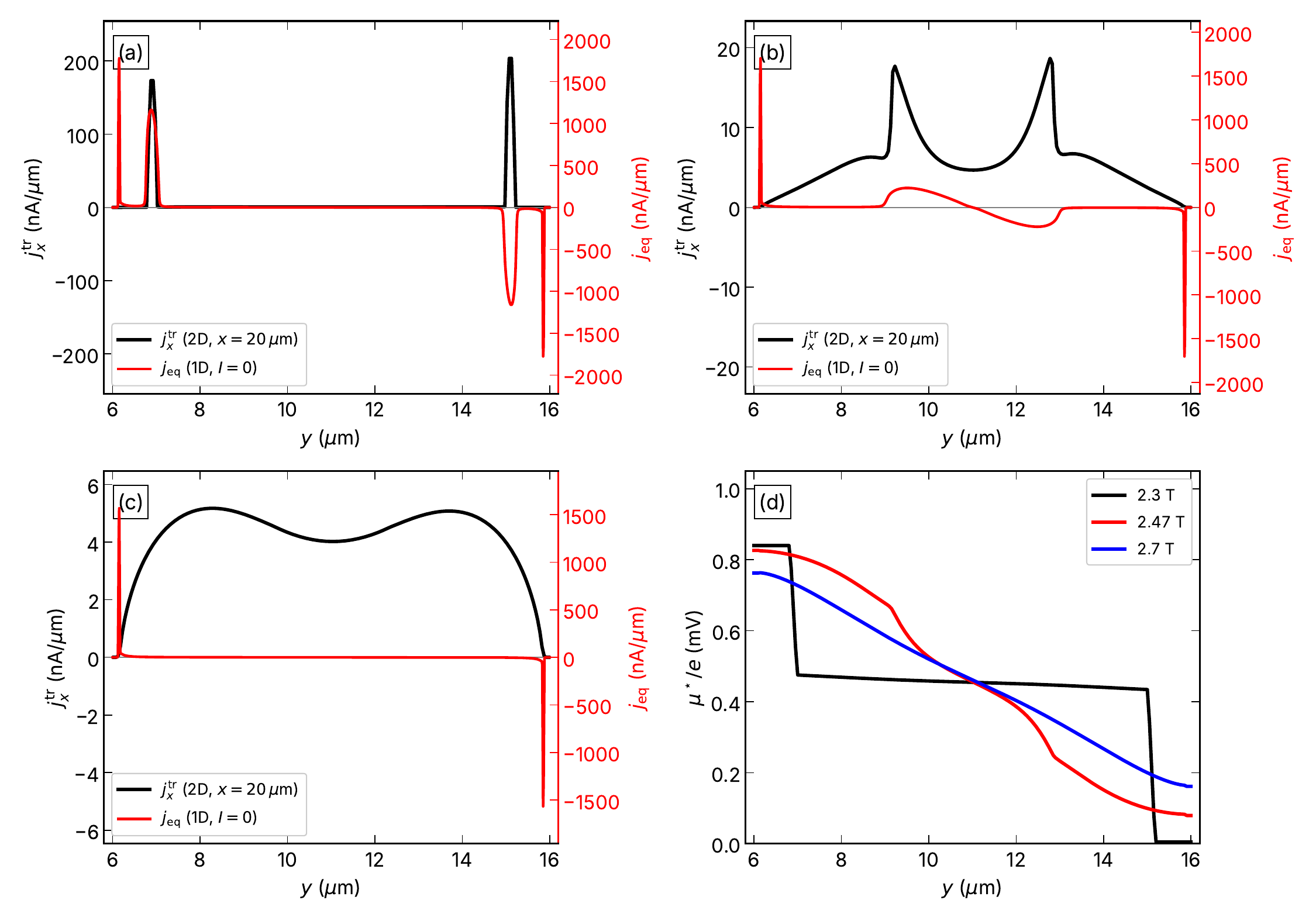}
\caption{Transport current density $j_x$ (black, left scale) across the bar at $x=20\um$, midway between the interior contacts, in the 3T configuration at (a)~$2.3$, (b)~$2.47$ and (c)~$2.7$~T, together with the equilibrium current density $j_{\rm eq}$ of the translation-invariant bar at the same fields (red, right scale). The transport current has the same sign across the whole cross section in all three regimes; the equilibrium current is opposite at the two edges. (d)~Electrochemical potential along the same cut; the Hall voltage drops across both edge strips at $2.3$~T, across the pinched bulk strip between the interior contacts at $2.47$~T, and nearly linearly at $2.7$~T.}
\label{fig:cuts}
\end{figure*}

Figures~\ref{fig:maps} and \ref{fig:cuts} show where the current flows in the device in the three regimes. At $2.3$~T the transport current follows the two edge strips, which wrap around the arms and run along the arm edges into the grounded contacts. The interior contacts are encircled by closed incompressible rings and carry no current, in agreement with the isolation of the interior contacts measured in this regime~\cite{Sirt2025APL}. At $2.47$~T the bulk strip is interrupted by the depleted rims of the interior contacts, and the current flows around them through the compressible rims, which are attached to the floating metal, before it enters the grounded arms through their compressible interiors. At $2.7$~T it spreads over the compressible bulk, divides around the interior contacts, and shows the classical hot spots at the diagonally opposite contact corners. In all three regimes the current density along a cut between the arms has the same sign across the full width [Fig.~\ref{fig:cuts}(a--c)]: the current emitted by contact 1 reaches contacts 3 and 5 along both edges of the bar, and the Hall voltage drops across both strips [Fig.~\ref{fig:cuts}(d)]. The terminal currents nevertheless obey Eq.~(\ref{eq:LB}) wherever $\sigl=0$ connects the contacts, because the potentials of the free edge segments, and only those, determine them. The equilibrium current, shown for comparison, is opposite at the two edges; it does not enter the balance of Eq.~(\ref{eq:IAB}) and no arrangement of contacts measures it.

\begin{figure*}
\includegraphics[width=0.8\textwidth]{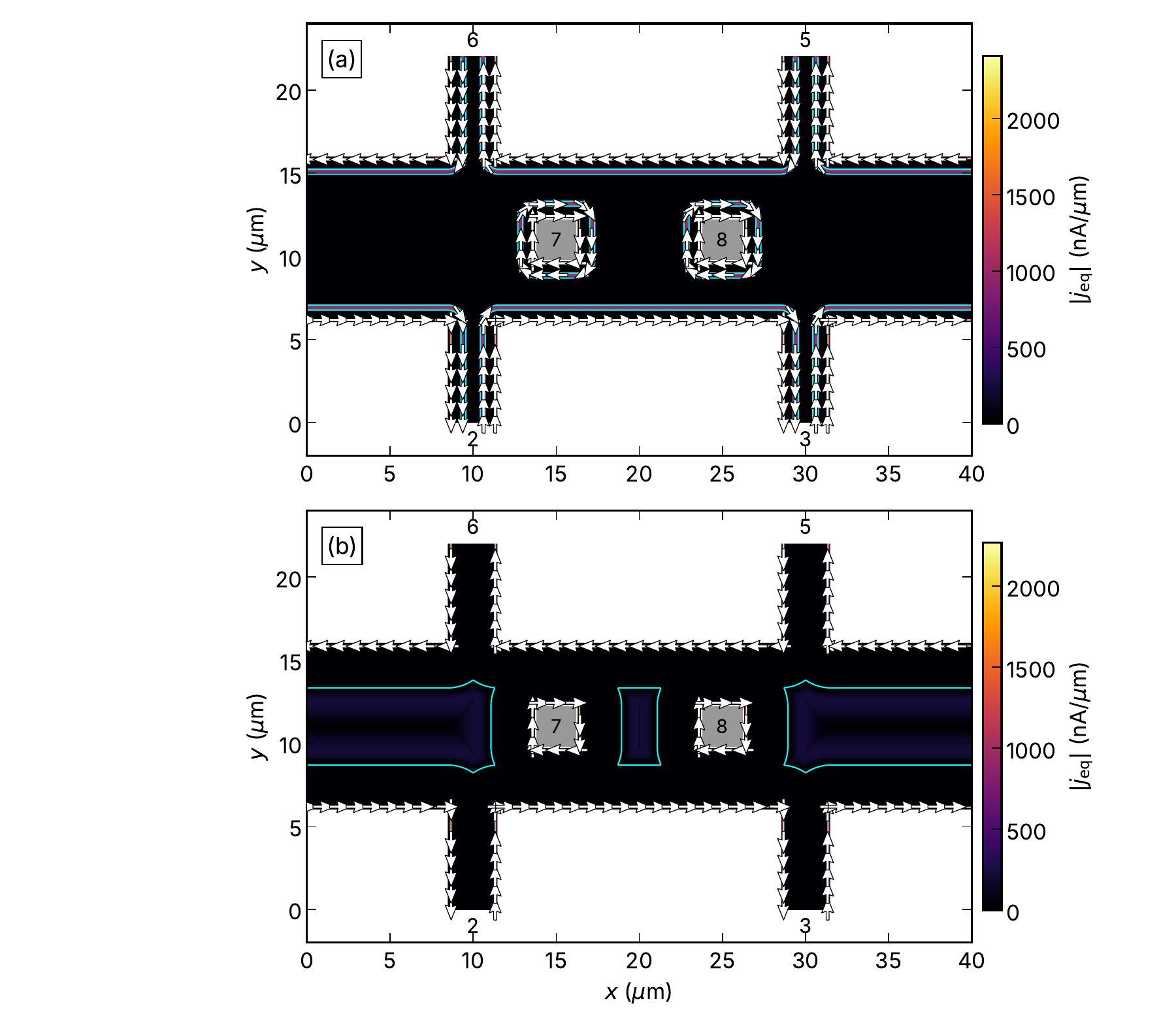}
\caption{Equilibrium current density (drift part) $\bm j_{\rm eq}=(e/h)\bar\nu\,\hat z\times\bm\nabla V$ in the device at (a)~$2.3$~T and (b)~$2.47$~T, from the self-consistent potential mapped by the distance to the nearest edge. Colour: $|j_{\rm eq}|$; arrows: direction; cyan lines: $\bar\nu=2$. The current circulates in one sense around the whole perimeter, including the arms, and in the opposite sense around each interior contact; its net flux through any cross section vanishes. In (a) it is concentrated in the edge strips and the depletion zones, in (b) a weaker part is spread over the bulk strip. The density-gradient current of the compressible regions (Fig.~\ref{fig:profiles}) is not included; it circulates in the opposite sense within each compressible region and reduces the integrated current of every edge region to the bulk magnetization without changing the pattern.}
\label{fig:eqmaps}
\end{figure*}

Figure~\ref{fig:eqmaps} shows this equilibrium current in the device. It flows in one sense around the entire perimeter, following the arms and the depletion zones, and in the opposite sense around each interior contact, as a magnetization current must; its flux through the cut of Fig.~\ref{fig:cuts} is zero to numerical precision. Figures~\ref{fig:maps} and \ref{fig:eqmaps} side by side are the content of the title: the chiral current is the one that carries nothing; the current that carries the $66$~nA emitted by contact $1$ at the $1$~mV bias of the partition experiment is unidirectional. (The $200$~nA of Table~\ref{tab:edges} are the nonlinear translation-invariant runs; the device calculation is linear response at the measured bias.)

\subsection{Floating probes, interior contacts and four-terminal resistances}
\label{sec:probes}

\begin{figure*}
\includegraphics[width=\textwidth]{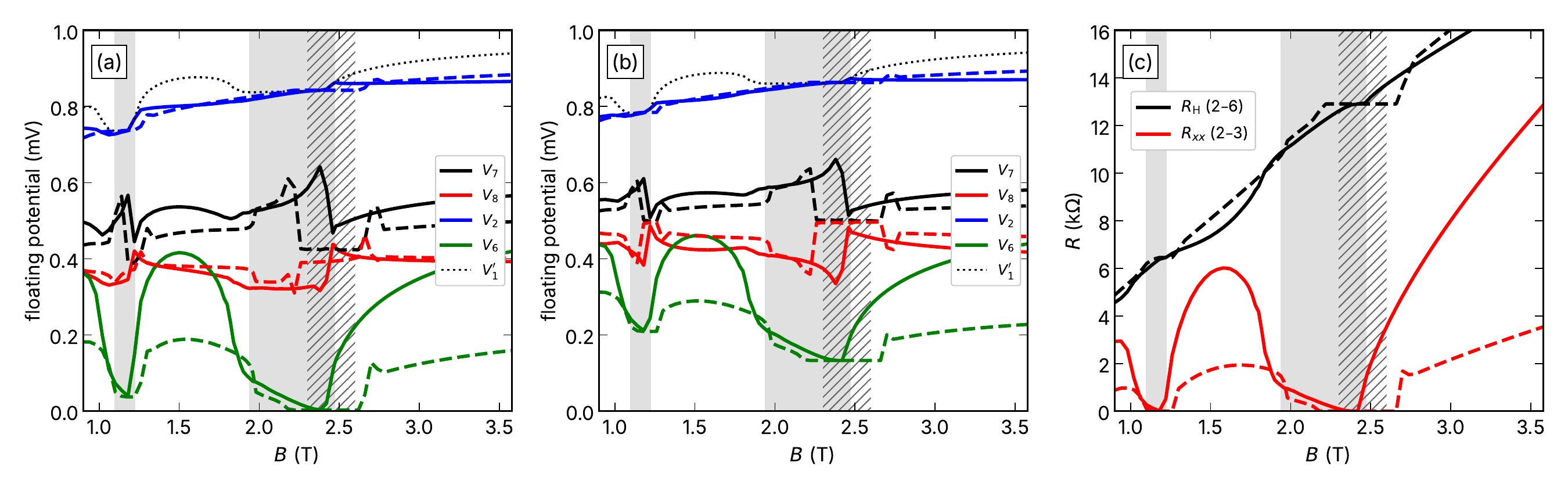}
\caption{Potentials of the floating contacts and four-terminal resistances from the two-dimensional solution, clean (solid) and with the disorder phenomenology (dashed). (a)~3T configuration: interior contacts $V_7$ (black) and $V_8$ (red), arm $V_2$ (blue), arm $V_6$ (green), and the potential $V_1'=V_1-I_1R_1$ behind the contact resistance of the biased contact (dotted). (b)~The same for the Hall configuration, current from contact 1 to contact 4 with all other contacts floating. (c)~Four-terminal $R_{\rm H}=|V_2-V_6|/I$ (black) and $R_{xx}=|V_2-V_3|/I$ (red) of the Hall configuration. Shaded: plateaus of the clean translation-invariant solution; hatched: the measured $\nu=2$ plateau.}
\label{fig:probes}
\end{figure*}

The two-dimensional solution gives the potentials of all floating contacts, and these contain predictions that the same sample can test. The sample was made for the interior-contact experiments~\cite{Kendirlik2017,Sirt2025APL}. In Ref.~\cite{Kendirlik2017} the two interior contacts were used as two-terminal probes of the bulk, a finite impedance between them on the plateau having been predicted~\cite{Yildiz2013}; in Ref.~\cite{Sirt2025APL} one interior contact was grounded and the current $I_A$ it collects was measured together with the peripheral voltages: $I_A=0$ over most of the plateau, and a finite $I_A$, with the corrected Hall resistance still quantized, on the high-field side of each plateau, which Ref.~\cite{Sirt2025APL} interprets with the screening theory as the current moving from the edge strips into the bulk. During the partition measurements the two interior contacts were present and floating. What is proposed here is different and has, to our knowledge, not been measured: their floating potentials, read against the current contacts, which can decide in a single field sweep what the partition cannot decide at all. Figure~\ref{fig:probes} shows them for the partition configuration and for the usual Hall configuration (current from 1 to 4). The arms behave as Eq.~(\ref{eq:LB}) requires: on the plateau the floating arm 2 takes the potential $V_1'$ of the contact upstream of it and the arm 6 that of contact 5 in the 3T configuration or of contact 4 in the Hall configuration, so that $V_2-V_6$ is the quantized Hall voltage; between the plateaus the dissipative bulk pulls $V_6$ up and $V_2$ down. This is the chiral propagation of the electrochemical potential in a local-Ohm solution without edge channels, and it is what makes the four-terminal resistances of Fig.~\ref{fig:probes}(c) come out as they do: with the disorder phenomenology $R_{\rm H}$ is quantized and $R_{xx}=0$ from $2.2$ to $2.66$~T, which covers the measured plateau, and $R_{xx}$ above it is $1.7$--$2.7$~k$\Omega$ between $2.7$ and $3.2$~T, against measured maxima of $2$~k$\Omega$ at $2.75$~T and $1.2$~k$\Omega$ at $3.2$~T~\cite{Sirt2025NJP}; the clean solution gives the narrower plateau of Fig.~\ref{fig:sweep}(c) and $5$--$10$~k$\Omega$ between the plateaus.

The interior contacts are more interesting. In the edge-strip regime of the clean solution the compressible bulk between the two strips is a dissipative conductor: it touches the end contacts $1$ and $4$ and the compressible interiors of the arms, carries a small dissipative current, and its potential falls along the bar from that of contact $1$ to that of contact $4$. The interior contacts, each isolated by its incompressible ring, read this bulk potential through the residual conductivity of the ring, and the two of them, $10\um$ apart, register different fractions of the Hall voltage: in the 3T configuration at $2.3$~T, $V_7=0.58$ and $V_8=0.33$~mV for $V_1'=0.84$~mV, and in the Hall configuration $0.62$ and $0.38$~mV for $V_1'=0.86$ and $V_4'=0.13$~mV. The difference $V_7-V_8$ grows from $0.19$--$0.22$~mV at $2.1$~T to $0.33$~mV at $2.38$~T, i.e.\ from a quarter to nearly half of the Hall voltage, as the strips move inward and the bulk between them narrows, and it collapses to $0.04$~mV at $2.47$~T, where the bulk becomes incompressible. Wherever the region around the interior contacts is quantized, in the clean solution at the merging field and in the disordered variant over the whole measured plateau, both contacts sit midway between the potentials of the two current contacts, $V_7=V_8=(V_1'+V_4')/2$ to within $0.01$~mV in the Hall configuration, the potential of the quantized bulk between the arms dropping linearly across the bar in the solution; in the 3T configuration the grounded arm $3$ shifts $V_8$ slightly below this midpoint. Within the model the difference $V_7-V_8$ is therefore a direct diagnostic of whether the bulk between the strips is compressible, i.e.\ whether the current is carried by edge strips or by the bulk, the difference and not the absolute values being the robust quantity (see below). This is the transition observed with these interior contacts in Refs.~\cite{Kendirlik2017,Sirt2025APL}, and it is the quantity on which the clean and the disordered variants of the model differ most: the clean bulk gives a difference of up to half the Hall voltage on the low-field side of the plateau, the localized bulk gives zero. A word on the mechanism that assigns these potentials, since a contact enclosed by a ring of $\sigl=0$ exchanges no current with its surroundings. The potential of an island coupled only weakly to its surroundings is the coupling-weighted average of the potential around it, whatever the magnitude of the coupling, as long as that coupling dominates the loading by the voltmeter. In the calculation the coupling is the residual conductivity of the ring, and the result is insensitive to its magnitude, as it must be: varying the floor between $10^{-7}$ and $10^{-4}\,e^2/h$ changes $V_7$ and $V_8$ by less than $0.01$~mV, because the bulk potential around a $4\um$ ring is nearly uniform on the scale of its variation along the bar. In the experiment the coupling is the thermally activated or tunnelling conductance of the ring, which is what makes the interior contacts ``isolated'' but not disconnected in Refs.~\cite{Kendirlik2017,Sirt2025APL}; the prediction holds when this conductance exceeds the voltmeter loading, which is the ordinary condition for reading any high-impedance probe, and it fails, by charging, if the island were disconnected altogether. The midpoint value in the quantized regime rests on the same residual conductivity, and on more: at $\sigl=0$ exactly the interior potential of a quantized region is undetermined (Sec.~\ref{sec:formal}); for a small residual conductivity that is uniform, as thermal activation across a uniform gap is, $\mus$ is harmonic, and in the long bar between the arms, with the two free edges at the potentials of the two current contacts, the harmonic solution is linear across the bar to the accuracy of the computed $V_7=V_8=(V_1'+V_4')/2$, the arms and the holes perturbing it only near them (the $3\%$ offset of $V_8$ in the 3T configuration). A residual conductivity that is not uniform shifts the absolute values in the quantized regime: with a floor of $10^{-5}\,e^2/h$ in the half of the device on one side of the bar axis and $10^{-7}$ on the other side, $V_7$ and $V_8$ move together from $0.50$ to $0.15$~mV in the Hall configuration ($0.86$~mV with the halves exchanged), while $V_7-V_8$ stays at $0.005$~mV; in the 3T configuration the difference stays below $0.07$~mV. In the edge-strip regime of the clean solution the same asymmetry changes $V_7$, $V_8$ and their difference by less than $0.005$~mV. The robust observable is therefore the difference: zero when the bulk around the contacts is quantized, a quarter to a half of the Hall voltage, depending on the field, when it is compressible. Loading by the voltmeter matters as well: the island's coupling to its surroundings is $10^{-8}$~S in the calculation (edge-strip regime, floor $10^{-6}\,e^2/h$) and $2.5\times10^{-10}$~S in the quantized regime (localized bulk), and an input impedance below ten times the inverse of this coupling pulls the reading towards ground; in the experiment the coupling is the measured isolation conductance of the interior contact~\cite{Kendirlik2017,Sirt2025APL}, and the reading requires an electrometer-grade input, not a lock-in input of $10$~M$\Omega$. In the edge-channel picture the potential of an interior contact that is not reached by an edge channel must likewise be supplied by an additional coupling; the difference is that the local-Ohm solution contains it.

The protocol of Ref.~\cite{Sirt2025APL}, interior contact $7$ grounded through its $2.4$~k$\Omega$ in the Hall configuration, can be computed as well. The clean solution gives $I_A/I_S\approx1\%$ below the plateau, the leakage through the dissipative bulk, decreasing along the low-field side of the plateau and vanishing (below $10^{-4}$) between $2.30$ and $2.42$~T, when the ring is fully formed and the bulk carries no current, and $0.8$--$1.0\%$ again from $2.46$~T on, where the clean bulk becomes compressible and $R_{\rm H}$ leaves the plateau; the localized bulk gives $I_A<10^{-5}I_S$ from $2.06$ to $2.66$~T and $10\%$ outside this range, and a residual hopping conductivity of $10^{-3}\,e^2/h$ in the localized regions raises this only to $I_A\le2\times10^{-3}I_S$ on the plateau, because the Hall current of a dissipationless region flows around an equipotential island. The isolation over most of the plateau is thus reproduced by both variants, the onset of bulk current at the merging field by the clean one. What neither reproduces is a finite $I_A$ coexisting with a quantized $R_{\rm H}$ over the high-field half of the plateau: in the model the current reaches the contact only where the bulk is dissipative, or at the merging field itself. The missing element is the structure of the ohmic-contact rim, the ``thin Schottky barrier'' of Ref.~\cite{Sirt2025APL}, which the etched-edge profile assumed here does not represent; the grounded-contact current is the measurement that calibrates the rim, and the floating potentials are the ones that locate the current. The bound $\mu_D<\mu_A\le\tfrac12(\mu_S+\mu_D)$ that Ref.~\cite{Sirt2025APL} gives for the grounded contact is consistent with the midpoint of the floating one.

\section{Discussion}
\label{sec:discussion}

The results can be summarized in one sentence, with the qualifications that follow: in the quantum Hall regime the electrochemical potential propagates chirally, the equilibrium current circulates chirally, and, in the self-consistent solution for the measured device, the transport current flows unidirectionally. The first statement is the content of Eq.~(\ref{eq:LB}) and is shared by every description with $\sigl=0$ and a definite sign of $\sigH$; it is what the multiterminal partition measures, and the measurements of Ref.~\cite{Sirt2025NJP} establish it cleanly, together with its breakdown by dissipation between the plateaus. The second statement is a property of the ground state and is unobservable in transport. The third is the statement of G\"uven and Gerhardts~\cite{Guven2003}; it holds in the self-consistent solution for the measured device in every cross section between the contacts, with the local transport model of Sec.~\ref{sec:model}, and it is what the SFM experiments~\cite{Ahlswede2001,Dahlem2010,Panos2014} see when they find the Hall voltage dropping across both edge strips or across the bulk.

It may help to separate three statements that Ref.~\cite{Sirt2025NJP} treats as one. \emph{Measured}: on the plateaus the terminal currents obey Eq.~(\ref{eq:LB}) with the measured $R_k$, and between the plateaus they do not. This shows that on the plateaus the terminal currents are those of a conductor with $\sigl=0$ along the current paths, to within the accuracy of the measurement, that $\sigl$ is finite between them, and gives the sign of $\sigH$; it does not fix $\sigl$ pointwise (Sec.~\ref{sec:formal}). \emph{Defined}: Ref.~\cite{Sirt2025NJP} calls the conductance matrix of Eq.~(\ref{eq:LB}) chiral current flow (their Sec.~III\,F). With this definition the measurement proves chiral flow by definition, and the same definition makes the unidirectional Hall term of their Drude model chiral as well. \emph{Assumed}: that the current density behind the matrix is the one of the Landauer--B\"uttiker derivation, confined to one edge between adjacent contacts. The first two statements support each other by construction; the third is an addition. It is the third statement that the title of Ref.~\cite{Sirt2025NJP} announces, while the measurement establishes the first; the third is the one that terminal currents cannot test (Sec.~\ref{sec:formal}), and the self-consistent solution places the bias-induced current on both edges (Sec.~\ref{sec:currents}).

\begin{table*}
\caption{Correspondence between the observations and explanations of Refs.~\cite{Sirt2025NJP,Sirt2024arXiv} and the unidirectional local-Ohm solution of this paper.}
\label{tab:corr}
\begin{ruledtabular}
\begin{tabular}{p{0.30\textwidth}p{0.52\textwidth}p{0.11\textwidth}}
Statement in Refs.~\cite{Sirt2025NJP,Sirt2024arXiv} & Unidirectional solution & Here \\
\hline
Plateau partition equals the chiral conductance matrix with the measured $R_k$, for all configurations (``proof of chiral current flow'') & Follows from $\sigl=0$ and quantized $\sigH$ for any current distribution; reproduced to $1\%$ with the clean strips and to $0.5\%$ with the localized bulk, with the current in the same direction in both strips & Secs.~\ref{sec:formal}, \ref{sec:partition}; Figs.~\ref{fig:equiv}, \ref{fig:ratios} \\
Partition reverses with the direction of $B$ & Sign of $\sigH$ selects the free edge that carries $V_1'$; the current density keeps its sign & Sec.~\ref{sec:partition} \\
Between the plateaus the Drude model fits with a mobility $\mu(B)$ two orders of magnitude below $\mu_0$ (``non-chiral diffusive transport'') & Dissipative compressible bulk; the fitted $\mu(B)$ is $\sigH/(\sigl B)$ of the partially filled Landau level, $5$--$9$~m$^2$/Vs from the measured $R_{xx}$, $1$--$13$~m$^2$/Vs computed & Sec.~\ref{sec:partition} \\
Suppression of the largest branch, strongest for 4T$+6$, then 4T$+4$, then 4T$+2$ $\approx$ 3T & Geometry of the dissipative paths; same ordering, magnitude set by $\sigl$ & Sec.~\ref{sec:partition}; Fig.~\ref{fig:ratios} \\
Quantized $R_{\rm H}$ and $R_{xx}=0$ from $2.3$ to $2.6$~T, $R_{xx}$ maxima of $1$--$2$~k$\Omega$ between the plateaus & Localized bulk; plateau width set by the arms and the interior contacts; $R_{xx}$ of $1.7$--$2.7$~k$\Omega$ between the plateaus (clean: $5$--$10$~k$\Omega$) & Sec.~\ref{sec:probes}; Fig.~\ref{fig:probes} \\
Total current $=$ persistent $+$ imposed; the persistent current decreases with $I$ [(iv) of Ref.~\cite{Sirt2024arXiv}] & Total edge currents $I_{\rm eq}+I_{\rm tr,high}$ and $I_{\rm eq}-I_{\rm tr,low}$ with $I_{\rm eq}$ unchanged to $0.3\%$; $I_{\rm pers}\equiv I_{\rm low}$ decreases by $I_{\rm tr,low}$ & Sec.~\ref{sec:currents}; Table~\ref{tab:edges} \\
Imposed current flows only on the high-potential edge [(v) of Ref.~\cite{Sirt2024arXiv}] & A representation, $\mu_0=\mu_D$ in Eq.~(\ref{eq:fhigh}); the bias-induced current flows in both strips, equally in linear response & Sec.~\ref{sec:currents} \\
Current confined to one edge between adjacent contacts; ``ballistic'' on the plateaus & Not reproduced, and not measurable by contacts; the Hall-potential profiles of SFM show both edges, and the interior contacts can test it & Secs.~\ref{sec:formal}, \ref{sec:probes} \\
\end{tabular}
\end{ruledtabular}
\end{table*}

Table~\ref{tab:corr} collects the observations and the explanations of Refs.~\cite{Sirt2025NJP,Sirt2024arXiv} next to their counterparts in the unidirectional solution. Each has one: quantitatively on the plateaus and for the ordering between them, semi-quantitatively for the magnitudes between the plateaus, where the clean calculation is off by a factor of $3$--$6$ in the dissipation and the three-number phenomenology brings it into range, and at the level of the total edge currents for the finite-current statements of Ref.~\cite{Sirt2024arXiv}. Two things the solution does not claim. It takes the quantized $\sigH$ of the strips as an input from the SCBA and does not resolve the microscopic origin of the quantization; and obtaining the conductance matrix without edge channels does not demonstrate their absence. Self-consistent quantum-electrostatic calculations for a narrow ballistic conductor recover Landauer--B\"uttiker channels at low field, compressible and incompressible strips at high field and a hybrid regime between them~\cite{Armagnat2020}, the two descriptions being limits of one problem rather than two theories; the distinction between the response current and the Fermi-surface current that underlies Sec.~\ref{sec:currents} was drawn by Komiyama and Hirai~\cite{Komiyama1996}. The single statement without a counterpart is the confinement of the bias-induced current to one edge between adjacent contacts, and it is, by Sec.~\ref{sec:formal}, the one statement that the terminal currents could not have tested.

The chiral edge-channel picture and the screening picture are not in conflict about the terminal currents, which is why the current partition cannot decide between them; they differ in where the current flows, over which width, and with how much dissipation. In the screening picture the ``edge channel'' is the incompressible strip: a region of finite and $B$-dependent width, which moves from the edge to the bulk across the plateau, whose two members carry the transport current in the same direction, and which can merge into a single bulk strip without any change in the terminal currents. The experiments that distinguish the pictures are therefore those that resolve position or nonlinearity: the Hall-potential imaging already cited; interior contacts, which measure the potential of the bulk directly and showed the transition from edge to bulk current across the plateau~\cite{Kendirlik2017,Sirt2025APL}; and the current-direction and bias asymmetries of the nonlinear response~\cite{Siddiki2009,Siddiki2010,Panos2014}, which follow from the self-consistent widening of the high-$\mus$ strip shown in Table~\ref{tab:edges} and have no counterpart in a picture of fixed one-dimensional channels.

Several limitations of the present calculation should be kept in mind. The primary calculation treats the sample as homogeneous, so that the plateaus end where the clean bulk becomes compressible: the translation-invariant plateau lies $0.3$~T below the measured one, the device plateau is narrower, and the dissipation between the plateaus is too large by a factor of $3$--$6$. The disorder phenomenology of Sec.~\ref{sec:model} shows that a localization window of $\pm0.2$ in the average filling factor, a $300$~nm smearing of the strips and a fourfold reduction of $\sigl$ recover the measured plateau width, the measured longitudinal resistance between the plateaus and the suppression of the partition there; but it is a parametrization of long-range disorder and spin splitting, not a theory of either, its three parameters are not independent, and the localization of a compressible bulk at $|\bar\nu-2|<0.2$ is exactly what the interior contacts can confirm or refute (Sec.~\ref{sec:probes}). The spin-split plateaus are omitted, so that the $\nu=1$ plateau, which Ref.~\cite{Sirt2025NJP} also fits with Eq.~(\ref{eq:LB}), is covered by Sec.~\ref{sec:formal} but not by the present SCBA calculation. It should also be said plainly, since this paper criticizes a fitted mobility, that the three numbers of the phenomenology are fixed once for all fields and all configurations, whereas the Drude description of Ref.~\cite{Sirt2025NJP} needs a free function $\mu(B)$; and that the primary, clean calculation has no adjustable parameter at all. The device calculation is a two-dimensional transport calculation with mapped one-dimensional self-consistent profiles, not a two-dimensional self-consistent electrostatic solution: the density in the arms, at the junctions and around the interior contacts is taken from the translation-invariant profile by the distance to the nearest edge, which is least controlled exactly where several boundaries are close, the depletion at the interior-contact rims is assumed equal to that of an etched edge, the contacts are treated as ideal equipotentials behind the measured series resistances, and the bias-induced displacement of the strips is not fed back. The equilibrium current maps show the drift part; the complete equilibrium current of an edge region, the bulk magnetization, is given in Sec.~\ref{sec:currents}, and the counter-flowing density-gradient currents of the compressible regions~\cite{Geller1994,Geller1995} are shown only for the translation-invariant bar. None of these simplifications touches the formal result of Sec.~\ref{sec:formal}, which is exact for $\sigl=0$.

\section{Conclusion}

Multiterminal current partition on a quantum Hall plateau measures the sign of the Hall conductivity and the absence of longitudinal dissipation. Each observation and each explanation of Refs.~\cite{Sirt2025NJP,Sirt2024arXiv} has a counterpart in the unidirectional local-Ohm solution (Table~\ref{tab:corr}); a test that every dissipationless Hall conductor passes does not single out edge channels. Terminal currents fix the net current through every cross section and nothing about the circulating part of the current density, and the chirality of the current density is a statement about the circulating part; it is therefore an assumption of the model used for the interpretation, not a result of the measurement. Any two-dimensional conductor with $\sigl=0$ and $\sigH=\nu e^2/h$ reproduces the ``coherent and chiral'' conductance matrix, and the self-consistent screening theory with a local Ohm's law reproduces the plateau partition of the measured device for all contact configurations, as well as the configuration-dependent suppression between the plateaus, without edge channels and without an adjustable mobility. In the self-consistent solution the current density consists of a chiral equilibrium current, which carries no net current and whose edge value is the bulk magnetization, and a transport current which, in the cross sections between the arms, flows in the same direction in both edge strips and in the bulk strip and carries the measured current; the only counter-flow of the transport current is in loops of a few percent around floating contacts. The imposed current of Ref.~\cite{Sirt2024arXiv} is confined to one edge only by the choice of the reference chemical potential; the bias-induced current is shared by both edges, equally in linear response and increasingly unequally with current, in agreement with scanning-force-microscopy experiments. Deciding where the current flows requires probes that resolve position or nonlinearity; the terminal currents of a Hall bar on a plateau do not. The sample of Ref.~\cite{Sirt2025NJP} carries such probes: read with an input impedance well above their isolation resistance, its two interior contacts should sit at a common potential wherever the bulk around them is quantized, and differ by up to half the Hall voltage wherever the current flows in edge strips around a compressible bulk. Reading them in the same cool-down as the partition would locate the current where the partition cannot.

\begin{acknowledgments}
The author thanks S.~Ludwig and S.~Sirt for many discussions on current flow in the quantum Hall regime over the years, and R.~R.~Gerhardts for the foundations on which this work rests. The large-language-model assistants Claude (Anthropic) and ChatGPT (OpenAI), as well as Jenni AI and Grammarly, were used to assist with the translation of the SCBLIN code from Fortran to Python, the scripting of the calculations and figures, and the English; the author takes full responsibility for the content.
\end{acknowledgments}

\end{document}